\documentclass[showpacs,amssymb,floatfix,pre,aps,notitlepage]{revtex4-1}
\usepackage[utf8]{inputenc}
\usepackage{amsmath}
\usepackage{amsfonts}
\usepackage{graphicx} 
\usepackage{amssymb}
\usepackage{placeins}
\usepackage{psfrag}
\usepackage{color}

\usepackage{array}
\usepackage{psfrag}
\usepackage{comment}
\newcommand{\beq}{\begin{equation}}
\newcommand{\eeq}{\end{equation}}
\def\beqa#1\eeqa{\begin{align}#1\end{align}}
\newcommand{\bv}{ \mathbf v }
\newcommand{\Dm}{ {\mathbb D_m}}
\newcommand{\Dmp}{ {\mathbb D_x}}
\newcommand{\RR}{\mathbb R}
\DeclareMathOperator*{\dt}{\partial_t }
\DeclareMathOperator{\Tr}{Tr }
\DeclareMathOperator{\Det}{Det }
\newcommand{\mcl}{\mathcal}
\begin{document}

\title{Polydispersity and optimal relaxation in the hard sphere fluid}
\author{Matthieu Barbier}
\author{Emmanuel Trizac}
\affiliation{Laboratoire de Physique Th\'eorique et Mod\`eles Statistiques, 
UMR CNRS 8626, Universit\'e Paris-Sud, 91405 Orsay, France}

\begin{abstract}
 We consider the mass heterogeneity in a gas of polydisperse hard particles as a key to optimizing a dynamical property: the kinetic relaxation rate.  Using the framework of the Boltzmann equation, we study the long time approach of a perturbed velocity distribution toward the equilibrium Maxwellian solution.  We work out the cases of discrete as well as continuous  distributions of masses, as found in dilute fluids of mesoscopic particles such as granular matter and colloids. On the basis of analytical and numerical evidence, we formulate 
a dynamical equipartition principle that leads to the result 
that no such continuous dispersion in fact minimizes the relaxation time, as the global optimum is characterized by a finite number of species.  This optimal mixture is found to depend on the dimension $d$ of space, ranging from five species for $d=1$ to a single one for $d\geq 4$. The role of the collisional kernel is also discussed, and extensions to dissipative systems are shown to be possible. 
\end{abstract}

\date{\today}

\maketitle

\section{Introduction}

Polydispersity is defined as the fact that the constituent particles of a system  are heterogeneous, exhibiting a spread in radius, mass, shape or some other material property. 
Its consequences on the dynamics of a gas with contact interactions have already been considered in a number of physical settings: both molecular gas mixtures and systems with mesoscopic components such as granular gases, colloidal suspensions and polymers which are almost inevitably polydisperse. Most often, elastic fluids are studied with the tools of equilibrium statistical mechanics, evidencing such properties as mixture fractionation \cite{Evans,WiSo10}, phase transitions \cite{ZBTC99,SeCu01} and local segregation \cite{Pagonabarraga}, while inelastic mixtures exhibit non-equilibrium effects such as granular segregation \cite{Ottino,Schroter,Garz06} 
and the like \cite{barrat2005,ArTs06}. 
The ensuing physics is non-trivial and constitutes
a theoretical challenge \cite{KrTa03}, as epitomized by the 
Soret effect and related transport phenomena in multi-component systems, involving both equilibrium and non-equilibrium mechanisms \cite{Bearman,Morozov,TrHa95}.

In the present work, we are interested in the dynamics of a slightly out-of-equilibrium homogeneous mixture, and specifically in characterizing the composition corresponding to the maximal relaxation rate toward equilibrium, as a first step toward optimization of dynamical properties such as transport coefficients. Since we focus largely on the one-dimensional case, where the radii of the particles are not relevant, their size and shape distribution is disregarded throughout this paper, and we characterize each species by their masses and concentrations only. 
A clear illustration of how a spread in mass may improve the relaxation rate of a gas is provided by the elementary case of the one-dimensional equimolar binary mixture. We consider two species in equal proportions, therefore retaining as the single parameter the mass ratio $\zeta = m_2/m_1$, values $\zeta$ and $1/\zeta$ being equivalent by symmetry.
The argument then goes as follows: the monodisperse limit $m_1 = m_2$ prevents equilibration, since 
in one dimension, the velocity distribution does not evolve as velocities are simply exchanged between same-mass particles upon collision. Likewise, in the infinite mass ratio $m_2/m_1\to \infty$ case, the heavy particles become reflecting walls for the light particles, and the system does not equilibrate either. 
Therefore, there must exist some optimal ratio in the interval $]1,\infty[$ over which the relaxation rate is always finite.
Following numerical observations for a binary mixture of hard rods \cite{Marro}, this problem has been solved for Maxwellian particles (see below) by two different techniques \cite{Mohazzabi,Mohaz2,Mohaz3,Dickman}, leading to the same result: an optimal mass ratio of 
$m_2/m_1 = 3 + 2 \sqrt 2\simeq 5.82$. This result has been extended to species with different concentrations in \cite{Eder86}, using a method that may be generalized to an arbitrary number of discrete species. 

Our main interest in this paper lies in the problem of the existence of an optimal continuous mass dispersion. We shall 
argue that no such dispersion may be found, as the fastest return to equilibrium will be obtained with a finite number of species. 
The grounds for this conjecture --which takes the form of a dynamical equipartition principle-- are exposed in section \ref{sec:cont}, after having first described in section \ref{sec:binary} the existing results and methods for a binary mixture that we build upon in our general treatment. We then investigate this optimum numerically by three different methods in section \ref{sec:num}. Finally, section \ref{sec:ext} presents extensions of the binary mixture problem to account for different collision rules, including inelasticity that translates the problem from a near-equilibrium setting to the vicinity of a non-equilibrium reference state: either a scaling solution or a non-equilibrium steady state if energy is injected into the system through external forcing.

\section{Elastic binary mixtures and beyond}
\label{sec:binary}

The problem of optimal relaxation in a one-dimensional binary gas mixture was first studied numerically in \cite{Marro} through molecular dynamics simulation of a system of hard rods. Investigations have subsequently concentrated on the homogeneous (space-independent) Boltzmann equation, describing the evolution of the velocity distributions of both species $F_i(v,t)$ ($i=1,2$) as a coupled system
\beq \partial_t \, F_i(v_i,t) = c_j \int \! dv_j\, |v_i-v_j|^\nu \left[\, F_i(v_i^*,t)F_j(v_j^*,t)- F_i(v_i,t)F_j(v_j,t) \, \right] \label{bolsys} \eeq
where $c_i$ is the concentration of the $i$ species ($c_1+c_2= 1$), 
the post-collisional velocities $v_i^*$ and $v_j^*$ are given by
\beqa v_i^* = (1-a_{ij})\, v_j + a_{ij}\, v_i \nonumber \\
 v_j^* = (1+a_{ij})\, v_i - a_{ij} \,v_j  \label{postcol} \eeqa
with $a_{ij}=(m_i-m_j)/(m_i+m_j)$, and the exponent $\nu$ represents the dependence of the collision frequency in the relative velocity of the particles. Hence, hard rods are associated with $\nu=1$, while we mostly focus here on Maxwellian rods with $\nu=0$, and briefly consider other choices in section \ref{sec:ext} such as very hard rods with $\nu=2$ \cite{Ernst81}. The latter two models have the advantage of allowing for easier analytical progress 
while still capturing the important physical effects of the
more realistic $\nu=1$ case \cite{GaSaBook}.
Note that while the Boltzmann equation in principle keeps track of collisions 
$i$-$i$ and $i$-$j$, the former do not contribute to the balance equation,
and a single collisional term therefore appears on the right-hand side of 
Eq. (\ref{bolsys}).

Two approaches were proposed to derive the optimal mass ratio: the first consists in taking the initial condition for $F_i(v,t)$ as a small perturbation around its equilibrium (Gaussian) state, so as to linearize the collision operator. This method employed in \cite{Dickman,Eder86} will serve as the basis for our own investigations. The second technique \cite{Mohazzabi,Mohaz2,Mohaz3} proceeds by discretizing time and choosing a specific initial condition, but retaining the full non-linear Boltzmann equation. It is reported in Appendix A, where we demonstrate how it can be connected to the former in the long-time limit. Both approaches thus lead to the same optimal mass ratio, that can also be 
recovered heuristically through the following argument \cite{Mohazzabi,Mohaz2,Mohaz3}. Given pre-collisional velocities $v_1$ and $v_2$ for particles of species 1 and 2, we are interested in relating the pre- and post-collisional difference in kinetic energy $\Delta E= ( m_1 v_1^{2} - m_2 v_2^{2})/2$ and $\Delta E^* =( m_1 v_1^{*2} - m_2 v_2^{*2})/2$. We define $\zeta=m_2/m_1$, and see from \eqref{postcol} that
\beq \Delta E^* =  \dfrac{4 \zeta (1-\zeta)}{(1+\zeta)^2} v_1 v_2 + \dfrac{1 - 6\zeta + \zeta^2}{(1+\zeta)^2} \Delta E. 
\eeq 
We next consider collisional averages $\langle...\rangle_c$, defined by averaging a given quantity over
consecutive collisions. It is a specific feature of Maxwell models that 
$\langle v_1 v_2\rangle_c = 0 $ under the assumption of isotropic velocity distributions, implying that
\beq\langle \Delta E^* \rangle_c = 
\dfrac{1 - 6 \zeta + \zeta^2}{(1+\zeta)^2} \Delta E
\eeq
We deduce that $\langle\Delta E^*\rangle_c$ vanishes for any value of $\Delta E$ if
$1 - 6 \zeta + \zeta^2=0$ and so
\beq \zeta=m_2/m_1 = 3 \pm 2\sqrt 2 . 
\label{stdratio} 
\eeq
This appears to yield a sufficient condition for optimal relaxation, from the fact that
the temperatures of each species equilibrate on average after one collision. 
Strictly speaking, there is no reason why a similar argument might hold
beyond the two species situation, and in addition, it is not even clear how then 
to generalize the heuristics.  In any case, arguments in this vein cannot in general yield
the solution to our problem, as they rely on several fragile assumptions, such as $\langle v_1 v_2\rangle_c = 0 $ which is not verified for non-Maxwellian models.


\subsection{State of the art}
\label{sub:lin}
The linearization approach involves the assumption that the system is close to equilibrium, and the velocity distribution $F_i$ can be written as a small perturbation around the Maxwellian equilibrium distribution $\mathcal M$:
\beq F_i(v,t)= c_i \mathcal M(m_i,v) \, (1+\psi_i(v,t)) \quad \hbox{with}\quad\mathcal M(m,v)=\sqrt{\dfrac{m}{2\pi}} e^{-mv^2/2}.\eeq
We may thus linearize Eq. \eqref{bolsys} for the perturbation
\beq \partial_t \, \psi_i(v_i,t) = c_j \int \! dv_j\, \mathcal M(m_i,v_i)\mathcal M(m_j,v_j)  \left[\, \psi_i(v_i^*,t) + \psi_j(v_j^*,t)- \psi_i(v_i,t) - \psi_j(v_j,t) \, \right] \label{bollin} \eeq
where we made use of the equality $\mathcal M(m_i,v_i^*)\mathcal M(m_j,v_j^*) = \mathcal M(m_i,v_i)\mathcal M(m_j,v_j)$ that arises from conservation of kinetic energy during a collision.
It proves convenient to expand the perturbation over the Hermite polynomials $H_n$ 
\beq\psi_i(v,t) = \sum_{n=1}^\infty U^n_i(t) \,H_n\left(v\sqrt {m_i/2}\right) \eeq
as they are orthogonal under the Gaussian weight given by the Maxwellian distribution
\beq \int_{-\infty}^{+\infty} dv \,\mathcal M(m,v)\, H_j(v\sqrt {m/2}) \,H_k(v\sqrt {m/2})=\sqrt{2\pi/m} \, 2^{k} \, k! \, \delta_{jk}, \label{ortho}\eeq
allowing us to relate the coefficient $U^n_i$ to the $n$th "Hermite moment" of the probability distribution $F_i$, and providing us with the convenient Mehler formula \cite[p. 198]{Rainville}
\beq
 \dfrac{1}{\sqrt{1-z^{2}}} \, \exp\!\left[-\left(\dfrac{x - yz}{\sqrt{1-z^{2}}} \right)^{2} \right] =  e^{-x^{2}} \sum_{n=0}^{\infty} \ \dfrac{z^n}{2^{n}\,n!} H_{n}(x) H_{n}(y)\ , \quad |z| < 1.
\label{Mehler}
\eeq
Performing changes of variable so as to integrate over $v_i^*$ and $v_j^*$ in the corresponding terms of Eq. \eqref{bollin} introduces displaced gaussians similar to the left-hand side of Eq. \eqref{Mehler}, with $z = a_{i j}$ and $z = \sqrt{1-a^{2}_{i j}}$. Using the orthogonality relation \eqref{ortho}, we may then rewrite the system \eqref{bollin} as simple algebraic equations coupling coefficients of the same order only:
\beq \dt \, U^{n}_i(t)=  c_j \left[ \, U^{n}_i(t) \, (a_{i j}^{n} -1) +  U^{n}_j(t) \,  (1-a_{ij}^2 )^{\frac{n}{2}} \right].\label{Udisc}
\eeq

\begin{figure}
\includegraphics*[width=190pt]{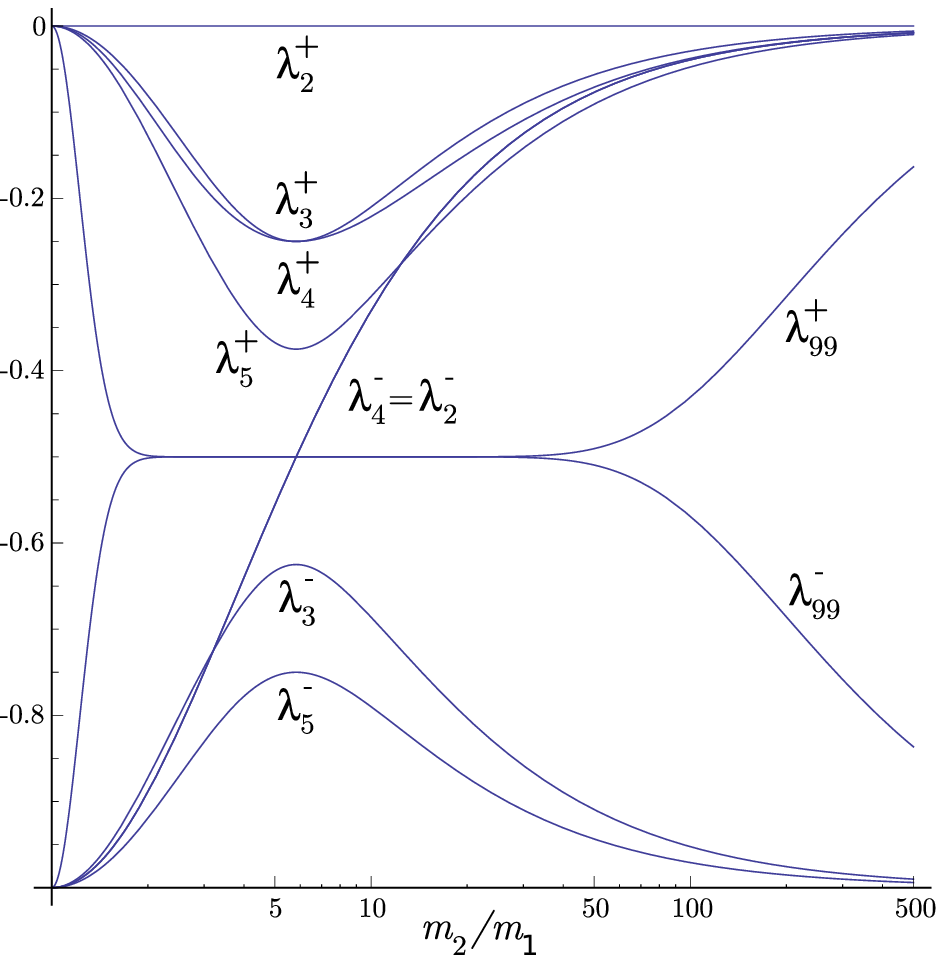}
\includegraphics*[width=190pt]{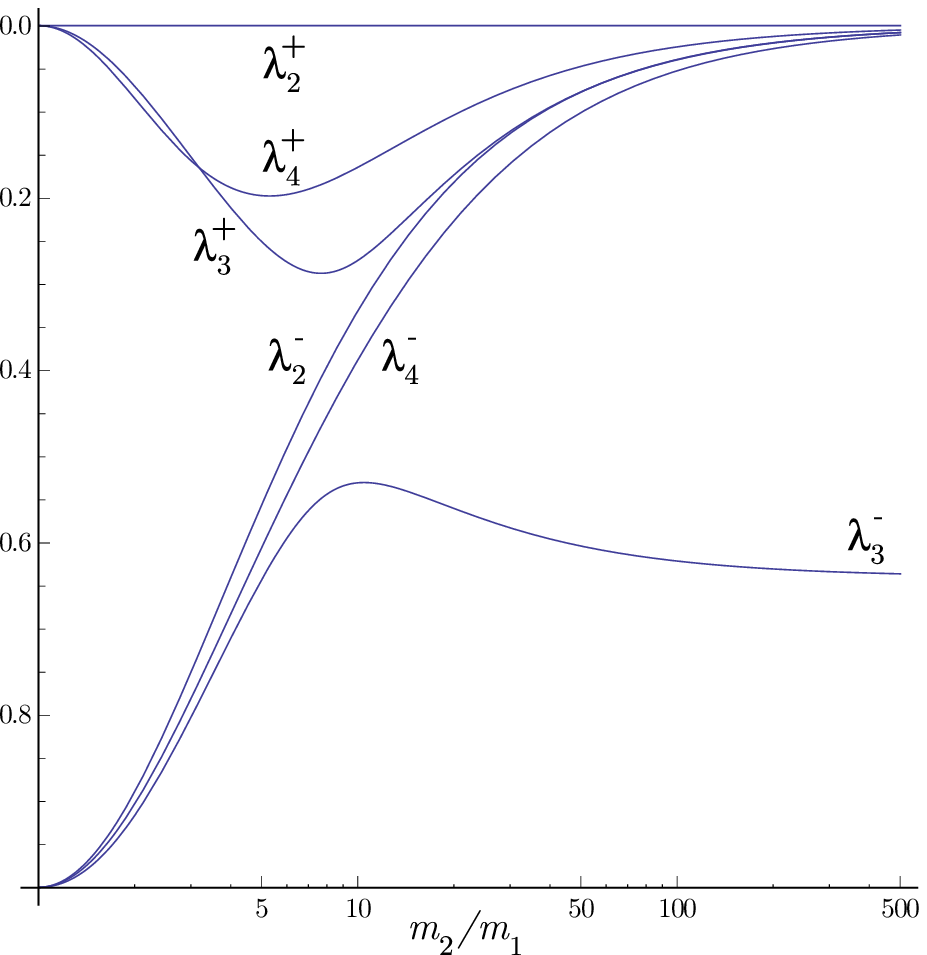}
\caption{Eigenvalue spectrum of the operator acting on $U^n$ on the right-hand side of Eq. (\ref{Udisc}), for a binary mixture, as a function of mass ratio
$\zeta = m_2/m_1$. Left: equimolar case $c_1 = 1/2$, all the extrema with $\zeta \neq 1$ are aligned at $\zeta=3+2\sqrt 2$; let us note that in this case $\lambda_2^-=\lambda_4^-=-4\zeta/(1+\zeta)^2$. Right: $c_1= 0.68$, the extrema are displaced and $\lambda_3^+$ and $\lambda_4^+$ now cross.}
\label{spectrum}
\end{figure} 

The whole spectrum of eigenvalues (see Fig.~\ref{spectrum}) for the time evolution of the $U^n_i$ was studied in \cite{Dickman} for an equimolar binary mixture $c_1 = c_2 = 1/2$. For each $n$, two non-positive eigenvalues are obtained
 \beq \lambda^{\pm}_n = \dfrac{1}{2}\left( \delta_e \,a_{12}^n-1 \pm \sqrt{\delta_o\, a_{12}^{2n} +(1- a_{12}^2)^n}\right) \leq 0
,\qquad \delta_e=1-\delta_o= \begin{cases} 1 & \text{if $n$ even} \\ 0 &\text{if $n$ odd}\end{cases}  \label{eigequi}
\eeq 
where by definition, $\lambda_n^+>\lambda_n^-$.
The highest eigenvalue vanishes for $n=1,2$ due to conservation of the total momentum and energy, associated with the first and second moments of $F_i$. The highest non-zero eigenvalue is associated with $n=3$ and is minimal for $m_1/m_2=3 \pm 2 \sqrt 2$, so that the longest lasting of all components in the perturbation $\psi_i$ appears to be related to the third moment and relaxes fastest for this precise mass ratio. The relevance of these eigenvalues to the relaxation rate of the system may be clarified by computing the Boltzmann $H$ function 
\beq H(t) = \sum_i \int \! dv F_i(v,t) \ln F_i(v,t)\eeq
using $\ln F_i(v,t) \approx \ln (c_i \mathcal M(m_i,v) ) + \psi_i(m,v)$
\beq H(t)- H(\infty)\approx \sum_i c_i \sqrt{\dfrac{2\pi}{ m_i}} \left( \sum_{n=1}^{\infty}n!\, 2^n\,  (U_i^n)^2 - 2 (U_i^2)^2  \right) \sim e^{2 \lambda^+ t} \qquad \hfill \left(\lambda^+ =  \max_{n>2} \lambda_n^+\right).\eeq

 Hence the highest non-zero eigenvalue in the complete moment hierarchy determines the long-time decay rate of the $H$ function, whose optimum is also given by \eqref{stdratio}. Here, all eigenvalues are found to be extremal for the same mass ratio if they have any local extremum on $\zeta \in ]1,\infty[$ but we will show that this property is not robust, and it should not be taken as a necessary condition for optimality: the optimal mixture  is  defined as the one that minimizes $\lambda^+$ and this definition remains appropriate in the more complex cases treated below. We also emphasize here that the $\lambda_n^-$ eigenvalues are irrelevant since there always exists a (non-zero) $+$ mode with slower relaxation: this is obvious for all $n>2$, and we always have found $\lambda_2^-$ to be dominated by other modes as well (see Fig.\ref{spectrum}-left where $\lambda_2^-$ coincides with $\lambda_4^-$, and also  Fig.\ref{spectrum}-right). 

\subsection{Non-equimolar binary and ternary mixtures}

\begin{figure}
\includegraphics*[width=180pt]{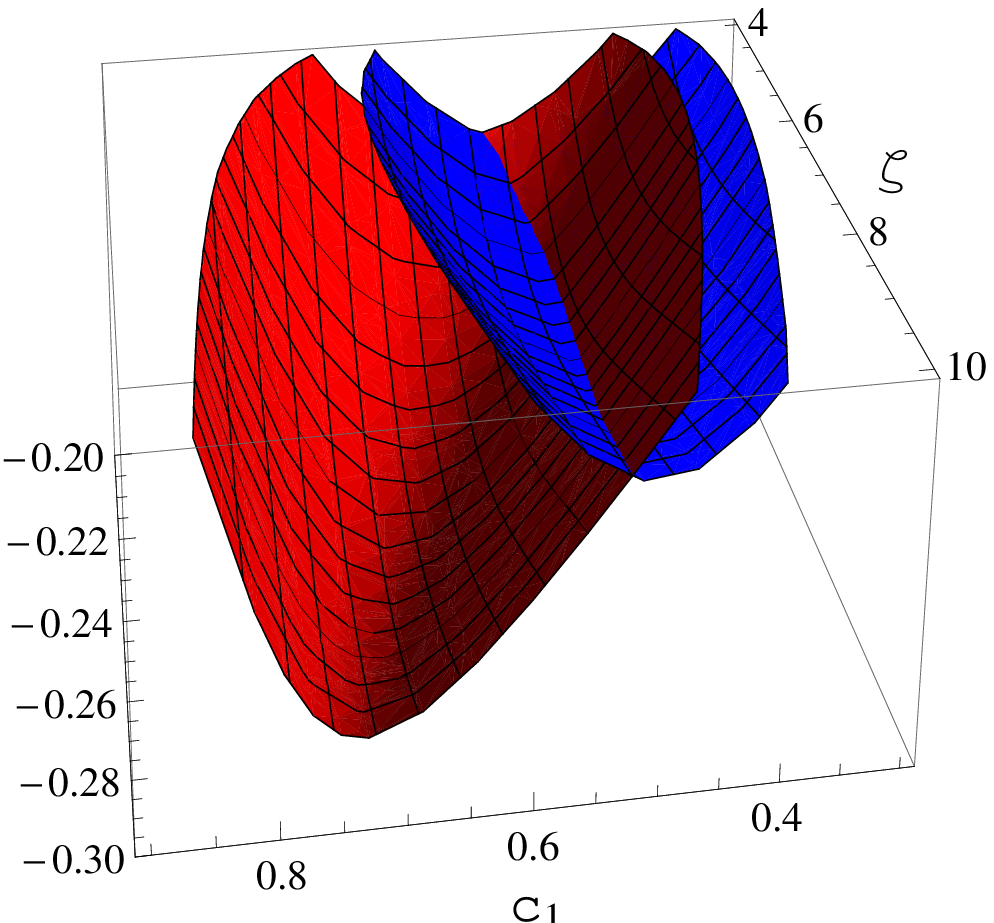}
\includegraphics*[width=130pt]{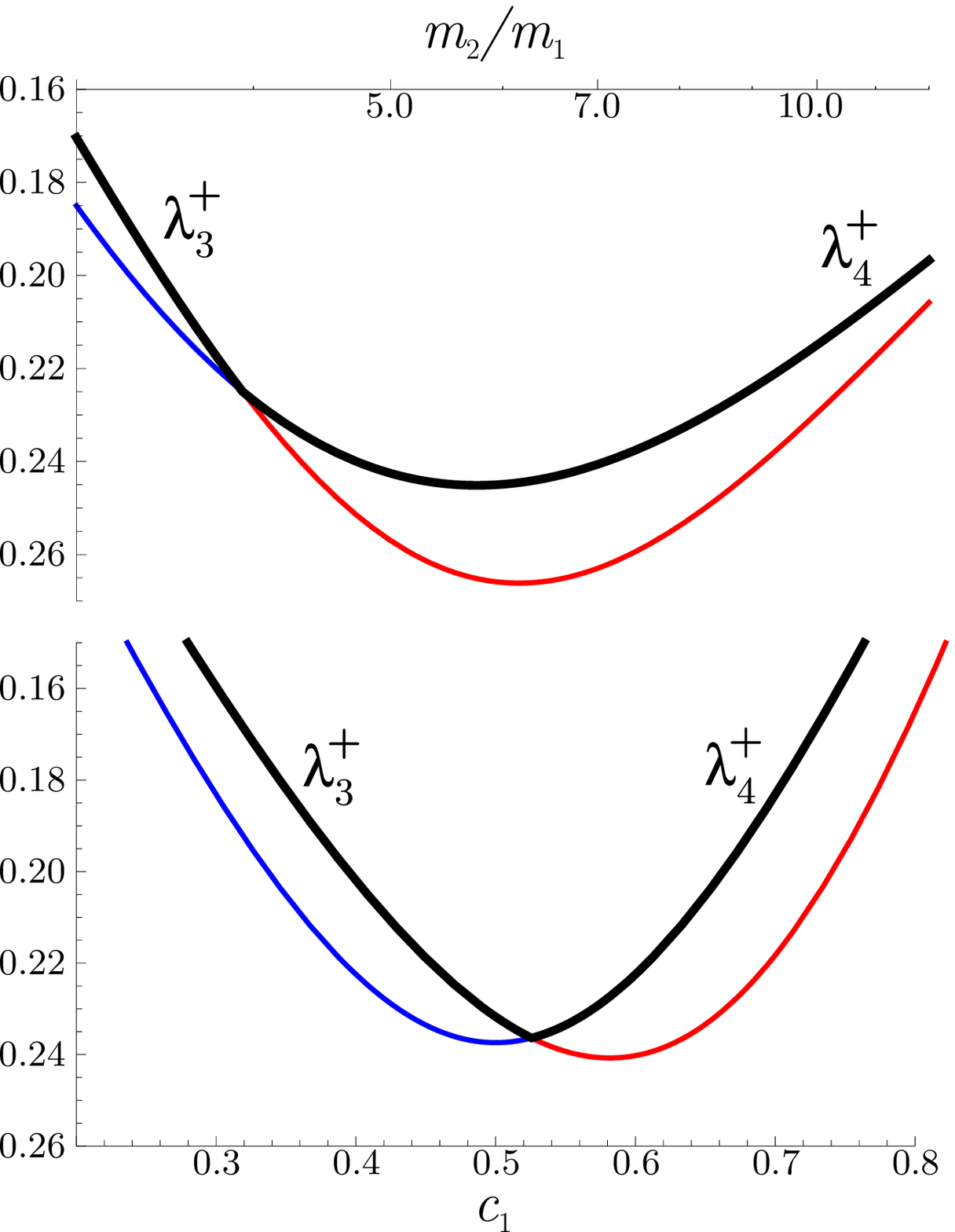}
\caption{Crossing of $\lambda_3^+$ and $\lambda_4^+$. Left: graphs of both eigenvalues against concentration $c_1$ and mass ratio $\zeta=m_2/m_1$. Right: section view of these surfaces for $c_1=0.6$ (top) and $\zeta = 4$ (bottom). The longest-living mode is highlighted in bold: it is the mode $n=3$ on the left of the intersections, and $n=4$ on the right. In principle, optimizing the relaxation involves finding the minimum of the composite curve (or surface) defined by the highest mode at each point.}
\label{eigcross}
\end{figure} 

The direct extensions of the previous results are now given for the non-equimolar binary mixture $c_1\neq c_2$, and the ternary mixture. No analysis of the optimum in these cases has been reported in the literature, although some aspects of the relaxation process for a specific initial condition are discussed in \cite{Eder86}. Yet they provide a test for the robustness of properties inferred from the elementary case above. We shall focus here on ascertaining which characteristics may be expected to hold in the general case, rather than present specific results exhaustively. In the non-equimolar binary mixture, the eigenvalues are given by
\beq \lambda^{\pm}_n = \dfrac{1}{2}\left( a_{12}^n(1-2c_1\delta_o)-1 \pm \sqrt{(1-2 c_1\delta_e)^2 (1-a_{12}^n-2c_1\delta_o)^2 +4 c_1 (c_1-1) (1- a_{12}^2)^n}\right)\eeq
where $\delta_e$ and $\delta_o$ are defined in \eqref{eigequi}. Let us also introduce the notation
\beq 
\lambda_n^* = \min_{ \lbrace m_i,c_i \rbrace} \lambda_n^+ \eeq
the minimum of the eigenvalue labeled $n$ over the whole parameter space. As $\lambda_3^+$ is always the highest eigenvalue in the equimolar case, our first guess would be to look for its general minimum $\lambda_3^*$; it is found for $c_1 \neq c_2$. The location of the optimum in terms of $c_1$ and $m_1/m_2$ is given by cumbersome analytical formulas, which may be evaluated to $c_1\approx 0.68$ and $m_2/m_1 \approx 7.7$, see Fig. \ref{spectrum}-right. 
However, $\lambda_3^+$ is actually not the highest eigenvalue at the point where it is minimum: a crossing appears between $\lambda^+_3$ and $\lambda^+_4$, as illustrated in Figs.~\ref{spectrum} and
\ref{eigcross}. Therefore, the largest eigenvalue is not associated with a given branch $n$: 
the longest living mode is either $n=3$ or $n=4$, depending on parameter values. In order to obtain the fastest relaxation for this mode, one must optimize the composite quantity defined as $\lambda^+=\max_{n>2} \lambda_n^+$ in each point, which reduces to $\max(\lambda_3^+,\lambda_4^+)$ if no other crossing is found. In other words, the optimization process
amounts to finding 
\beq 
\lambda^* = \min_{ \lbrace m_i,c_i \rbrace} \lambda^+  = \min_{ \lbrace m_i,c_i \rbrace} \max_{n} \lambda_n^+.  \label{eq:deflambdastar}
\eeq

Fortuitously, for a binary mixture, the absolute minimum of the composite quantity $\lambda^+$ is simply that of $\lambda_4^+$, which is equimolar and located at the mass ratio  \eqref{stdratio}. But beyond the binary case, it could be located on either mode or lie on their intersection, and is generically not equimolar. This can be appreciated in Table~\ref{tabtern} in the case of a ternary mixture. 
This discrepancy between the composite optimum $\lambda^* = \min(\lambda^+)$ and the moment optima $\lambda_n^*$ may be avoided by taking an initial condition that has a vanishing third Hermite moment (for instance a symmetrical velocity distribution), so that the eigenvalue to optimize is always $\lambda^+_4$, as it can be shown to cross no other eigenvalue over the whole parameter space.

\begin{table}[h]
\begin{tabular}{|c|c|c|c|c|c|}
\hline \textbf{Minimized quantity}& \textbf{ Value} & $m_1/m_2$ & $m_2/m_3$ &$ c_1 $&$ c_3$\\ 
\hline $\lambda_3^+$ & -0.3070 &0.226  & 0.226 & 0.517 &0.149\\ 
$\lambda_4^+$ &-0.2844 &0.226 & 0.226 & 0.296& 0.296\\
$\lambda^+$ & -0.2841 &0.235 & 0.232 &0.311 &0.286 \\
\hline 
\end{tabular} 
\caption{ The ternary mixture: analytical minimization of the eigenvalues $\lambda_3^+$ and $\lambda_4^+$, and numerical minimization of the composite value $\lambda^+= \max(\lambda_3^+,\lambda_4^+)$, over the masses $m_i$ and concentrations $c_i$. Parameters not specified in the table are fixed by the constraints of setting the sum of concentrations.  It is clear that the composite optimum lies on the intersection of the eigenvalues rather than on either one since it does not overlap with either $\lambda_3^*$ or $\lambda_4^*$ (scenario of Fig. \ref{eigcross}-right), despite being very close to the latter. We remark that optimal masses appear to be the same for $n=3$ and $n=4$ (although with different concentrations) and log-symmetric, although this is not the case for the composite optimum -- which moves even further away from this symmetry for a quaternary mixture. The symmetries appearing in these optimal distributions are discussed in section \ref{sub:opt}.}
\label{tabtern}
  \end{table}

 Extension of the present analysis to an arbitrary number $N$ of discrete species is straightforward, summing the right-hand members in Eqs. \eqref{bollin} and \eqref{Udisc} over all $j \neq i$.  Computation of the optimum cannot be performed analytically for $N>3$, and we detail in section \ref{sec:num} the methods and results of our numerical investigation: the corresponding optimum is in general non equimolar, and made up of only five species (in the one-dimensional system), irrespective of the starting value of $N\geq 5$.  We first intend to support this finite $N$ analysis by showing in the following section that such a  mixture should indeed be the generic optimum, even when initially assuming $N\to \infty$ or a continuous mass distribution.

\section{Continuous mass distributions}
\label{sec:cont}
\subsection{Operator and symmetries}
\label{sub:opt}
We consider the possibility of a continuous optimal mass distribution, instead of a mixture of discrete species. For the sake of simplicity, we confine ourselves to one-dimensional systems in this analysis, although it should not be invalidated in higher dimensions as demonstrated in section \ref{sec:num}. The extension of the approach detailed in section \ref{sub:lin} to a continuous dispersion is formally straightforward: we now define the probability distribution $F(m,v,t)$ as a function of mass, velocity and time, with mass distribution $c(m) = \int dv \,F(m,v,t)$, normalized to unity. This quantity
obeys the Boltzmann equation
\beq\dfrac{\partial F(m_1,v,t)}{\partial t}= \int_{\Dm} \! dm_2 \int\!\!\!\!\int_{-\infty}^{+\infty} \!dv_1 dv_2 \,  |v_1-v_2|^\nu \, F(m_1,v_1,t)\,F(m_2,v_2,t)\left[\delta(v_1^*-v) - \delta(v_1-v) \right] 
\eeq
with
\beqa &v_1^* = a(m_1,m_2) \,v_1-(a(m_1,m_2)-1)\,v_2 \\
&a(m_1,m_2)=(m_1-m_2)/(m_1+m_2).\eeqa
We have here defined the domain $\Dm$ as the support of the mass distribution. 
All functions of $m$ will be understood as defined over $\Dm$ (as any other value of the mass corresponds to a species that is absent from the system) which may in principle span all $\RR^+$.
We subsequently linearize $F$ around the equilibrium state $c(m) \mathcal M(m,v)$:
\beq F(m,v,t)= c(m) \mathcal M(m,v) \left(1 + \sum_{n=1}^\infty U_n(m,t) \,H_n\left(v\sqrt{ m/2}\right) \right). \label{eq:defpert}
\eeq
Eq. \eqref{Udisc} then becomes
\beq\dt \, U_n(m,t)= \int_\Dm \! dm' \ c(m') \left[ \, U_n(m,t) \, \left(a^{n}(m,m') -1\right) +  U_n(m',t) \,  \left(1 - a^2(m,m')\right)^{n/2} \right] 
\label{eq:Ucont}
\eeq

The perturbation $U_n(m,t)$ may be decomposed into eigenvectors of this integral operator, hereafter noted $\mcl K_n$:
\beqa & U_n(m,t) = \sum_i u^{(i)}_n(m)\,\, e^{  \lambda^{(i)}_n \,t} \\
& \lambda^{(i)}_n \, u^{(i)}_n(m)= \mcl K_n u^{(i)}_n(m) \eeqa
with  $\lambda_n^{(1)}> \lambda_n^{(2)}>...$ so as to differentiate between the eigenspaces. The $\lambda^{(i)}_n[c]$, $u^{(i)}_n[c]$ and $\mcl K_n[c]$ all depend functionally on the mass distribution $c(m)$, as will be explicited whenever necessary.

 We intend to find the dispersion $c(m)$ that optimizes the relaxation time of the whole system, thus we must extract the highest non-zero eigenvalue among all the $n$-indiced subsystems, then minimize it as a functional of $c$. Let us remark on the existence of these eigenvalues: the modified operator $\mathcal{C}^{1/2} \mcl K_n \mathcal{C}^{-1/2}$, where we define $\mathcal{C} \Phi(x) = c(x) \Phi(x)$, is easily seen to be symmetrical and a Hilbert-Schmidt operator as its kernel is square-integrable. The spectral theorem then applies to this operator, which has the same eigenvalues as $\mcl K_n$ with the modified eigenvectors $c(m) \,u^{(i)}_n(m)$. 

Finally, we should remark that the operator $\mcl K_n$ presents two useful symmetries. Firstly, it is invariant under a global rescaling $m\to \eta \, m$, as only mass ratios intervene as dimensionless quantities in the kernel. We may therefore always choose a mass distribution where the mean mass is fixed to unity. Secondly, in the case of even $n$ in which we are interested, we also see that the operator is invariant under inversion of all masses $m\to 1/m$, corresponding to changing $a(m,m')$ into $-a(m,m')$. Therefore, if $c(m)$ is optimal, then the same distribution with inverted masses is also optimal. In the case where there is only one optimum, it must be symmetrical with respect to $\ln (m)$.  This simple 
remark indicates that odd and even modes do not share the same $m$-symmetry:
indeed, the change $a \to -a$ leaves Eq. (\ref{eq:Ucont}) invariant for even values of $n$ only.

\subsection{Optimality condition}

The highest eigenvalue for a given order $\lambda^{(1)}_n$ (corresponding to the previously defined $\lambda_n^+$ in the binary mixture)   is given by the maximum of the Rayleigh quotient for operator $\mcl K_n$
\beq\lambda^{(1)}_n[c] = \max_{u} \dfrac{\langle \mcl K_n[c]\, u, u\rangle}{\langle u,u\rangle}\eeq
where we define the scalar product
\beq\langle f,g\rangle = \int_\Dm \! dm \,f(m)\, g(m) \,c(m)\eeq
including $c(m)$ in the measure in order to symmetrize the action of operator $\mcl K_n$ (as an alternative to studying the modified operator discussed in section \ref{sub:opt}) . We are first interested in minimizing $\lambda^{(1)}_n$ over $c$ in the space of $L_1$ positive functions, that is, find $c=c^*_n$ such that
\beq \lambda^{(1)}_n[c^*_n] = \min_{c\geq 0, \,c \in L_1} \lambda^{(1)}_n[c] = \lambda_n^*\eeq
where $\lambda_n^*$ is the absolute optimum, i.e. the minimum of the maximal eigenvalue for a given $n$.
This may also be understood as finding the pair of functions $(u^*_n, c^*_n)$ such that the Rayleigh quotient of $\mcl K_n$ is maximal over $u$ and minimal over $c$, i.e. a saddle point in the functional space $(u,c)$.

As the operator $\mcl K_n$ is linear in $c$, we may readily verify that $\lambda^{(1)}_n[c]$ is convex: indeed, if we perform a linear combination of distribution functions $c$ and $c'$, we see that
\beqa \lambda^{(1)}_n[\,\theta \, c + (1-\theta) \,c'] &=\max_{u} \left(\theta \dfrac{\langle \mcl K_n[ c] u, u\rangle}{\langle u,u\rangle} +  (1-\theta)  \dfrac{\langle \mcl K_n[c'] u, u\rangle}{\langle u,u\rangle} \right) \nonumber\\
& \leq \theta \max_{u} \dfrac{\langle \mcl K_n[ c] u, u\rangle}{\langle u,u\rangle} +  (1-\theta) \max_{u} \dfrac{\langle \mcl K_n[c'] u, u\rangle}{\langle u,u\rangle}. \eeqa
Therefore,
\beq \lambda^{(1)}_n[\theta c + (1-\theta) c'] \leq\theta \lambda^{(1)}_n[c] + (1-\theta) \lambda^{(1)}_n[c'] \eeq
verifies the condition for convexity, and it can be posited 
that any local minimum is also a global minimum.

We now come to our central conjecture, the underpinnings of which are discussed in 
appendix B. Based on numerical and analytical evidence, the surmise is that the desired saddle point for even $n$ is given by the couple $(u^*_n,c^*_n)$ verifying
\beqa  &  u^*_n(m)=1 \nonumber\\
& \lambda_n^* = \int_\Dm \! dm' \,c^*_n(m') \,j(m/m') \nonumber  \\
& j(m)= \left(\dfrac{m-1}{m+1}\right)^{n} -1 +  \,  \left(1-\left(\dfrac{m-1}{m+1}\right)^2 \right)^{\frac{n}{2}}.   
\label{saddle}
\eeqa
 The condition $ u^*_n(m)=1$ that the last remaining component of the perturbation $U_n(m)$ should have the same amplitude for all masses $m$ (for any Hermite moment of even degree $n$) 
can be viewed as a dynamical equipartition principle, as clarified and illustrated along our numerical results in section \ref{sec:num}. 
Note that the above expressions are not sufficient to determine univocally $c^*_n(m)$, as $\lambda_n^*$ is also unknown and must be found by minimization over positive normalized distributions; yet these equations allow us to deduce some essential properties of $c^*_n(m)$ in the next section. We must also compare the eigenvalues of subsystems associated with different degrees $n$. We first observe that, for even degrees, $\lambda_{2(n+1)}^{(1)}[c] < \lambda_{2n}^{(1)}[c]$ for all $c$, as both terms inside the integral
\beq 1+ \lambda_{2n}^{(1)}=\int_\Dm \! dm' \,c(m')\left[ a^{2n}(m,m') +  \,  \left(1-a^{2}(m,m') \right)^{n} \right]\eeq
are positive and decrease with increasing $n$.  
 We emphasize that we have not been able to find the counterpart of the principle (\ref{saddle}) 
for odd $n$. Yet, the corresponding eigenfunctions nevertheless share some features of $u_n^*(m)$ such as
non summability and positiveness, see section \ref{sec:num}, and these features seem sufficient to ensure many similarities between even and odd moments. Nevertheless, lack of definite analytical expressions in the latter case encourages us to focus on even moments first. Furthermore, eigenvalues associated with moments of different parity may cross, as already seen in section \ref{sub:lin} with a binary mixture.  For these reasons, it is simpler to consider a symmetric velocity distribution as initial condition, so that we may avoid taking odd moments into consideration. As for the non-zero eigenvalues associated with $n=2$, we will have to check whether any of them is higher than $\lambda_4^*$ for the same dispersion $c^*_4(m)$ once it is found.

\subsection{Existence of a continuous optimal dispersion}
\label{sub:disp}
We are interested in properties of the optimal mass distribution $c^*(m)$ defined as the solution of the second equation in \eqref{saddle}, which is a Fredholm integral equation of the first kind. We shall now see that no such solution exists if $\Dm$ is unbounded. First, we may convert Eq. \eqref{saddle} into a convolution equation over $\Dmp$
\beqa & x-y = \ln m \quad y = \ln m' \quad C(x)=e^x c^*_n(e^x)\nonumber \\ & \quad J(x)= j(e^x) = \tanh^n \left(\dfrac{x}{2}\right) -1 + \cosh^{-n}\left(\dfrac{x}{2}\right) \nonumber\\ & \lambda_n^* = \int_\Dmp \! dy \,C(y)\,J(x-y) \label{convo} \eeqa
where $\Dmp$ is the corresponding integration domain ($\Dm = \RR^+ \Rightarrow \Dmp = \RR$). Now it is easy to see that $ C(x) \in L_1(\Dmp)$ since
\beq\int_\Dmp dx |C(x)| = \int_\Dm \dfrac{dm}{m}\, |m \,c^*_n(m)| = \int_\Dm dm\, | c^*_n(m)| \quad \text{as }\Dm \subseteq \RR^+\eeq
and $c^*_n(m) \in L_1(\Dm)$ (by definition). Likewise,
\beq\int_\Dmp dx |J(x)| = \int_\Dm \dfrac{dm}{m}\, |j(m)| \eeq
and this integral is convergent as $j(m)$ is analytic on $\RR^+$ and verifies
\beq j(m) \approx \begin{cases}-2nm & \text{if } m \to 0 \\ -2n/m & \text{if } m\to \infty\end{cases} \eeq
hence $ J(x) \in L_1(\RR) \supseteq L_1(\Dmp)$. Now if $\Dmp = \RR$, Young's inequality for convolutions states
\beq||C \ast J ||_{1} \leq || C ||_1 \, ||J ||_1 \eeq
while here $C\ast J = \lambda^*_n \notin L_1(\RR)$. As a result, the domain $\Dmp$ must be bounded and $\Dm = [m_-, m_+]$. 

We may treat the integral \eqref{convo} as a convolution over $\RR$ with $C(y)=0$ for all 
$y \notin \Dmp$. However, this convolution $(C \ast J) (x)$ needs be equal to $\lambda_n^*$ only if $x \in \Dmp$, and may take any value outside of that domain. Accordingly, either $\Dmp$ contains non-empty intervals and $(C \ast J)(x)$ is constant inside these intervals and null or decreasing outside --entailing that it or its derivative is not continuous everywhere--  or $\Dmp$ is restricted to a set of points, the intersections of $(C\ast J)(x)$ with the constant function $\lambda_n^*$.

Since both $C(x)$ and $J(x)$ are in $L_1(\RR)$, there is no problem defining their Fourier transforms
\beqa \mathcal F[C](k)= \widehat C(k) = \int_\RR dx\,e^{-ikx} C(x) & & F[J](k) = \widehat J(k) = \int_\RR dx\,e^{-ikx} J(x)\eeqa
for which we have 
$\mathcal F[C*J](k) = \widehat C(k) \widehat J(k)$ so that we may write the convolution product as the inverse transform
\beq (C*J)(x) = \int_\RR \dfrac{dk}{2\pi} \,e^{ikx} \widehat C(k) \widehat J(k).\label{invft}\eeq
Due to the dominated convergence theorem, this product is everywhere continuous if it is possible to find an integrable function $g(k)$ such that
\beq|\,e^{ikx} \widehat C(k) \widehat J(k) |  \leq g(k). \eeq
However $\widehat C(k)$ is the Fourier transform of a $L_1$ function defined on a bounded interval, and it must be $L_\infty$ with
\beq C_0 \equiv ||\widehat C(k) ||_\infty \leq || C(x) ||_1.\eeq
The integrand in \eqref{invft} is dominated by $C_0 |\widehat J(k)|$, which is always integrable as we may easily show that both $J(x)$ and $\widehat J(k)$ have exponentially decreasing tails for any $n$. Both $||\widehat J(k)||_1$ and $||k \widehat J(k)||_1$ are finite, implying that $(C*J)(x)$ is both continuous and derivable everywhere. As a result, it is impossible to have $(C*J)(x) \in L_1(\RR)$ yet constant on non-empty intervals, and the domain $\Dmp$ must be restricted to a set of discrete points.

We have thus shown that no piecewise continuous function $c^*_n(m)$ may be found that verifies \eqref{saddle}, and hence our dynamical equipartition principle implies that the optimal dispersion is a discrete mixture rather than a continuum\footnote{We expect that, by a similar reasoning, $c^*_n(m)$ cannot be continuous given the more general condition of $u_n^*(m)$ being non-integrable and of constant sign, as seems to be relevant for odd values of $n$, see e.g. Fig. \ref{eigsys}. This eigenfunction might in principle be integrable, but it would require a fast decrease for large masses, and hence imply that extremely massive particles do not contribute in the slowest relaxation mode, whereas one expects the contrary on physical grounds.}. A heuristic understanding of this discrete optimum may be found in the idea that two species with very similar or very different masses cannot equilibrate, therefore optimal relaxation imposes some distance constraint on the masses of each pair of species, and these constraints cease to be satisfiable for an infinite number of infinitely close species. The value of these arguments becomes more tangible in the next section, where we observe that the optimal dispersion $c^*_n(m)$ not only reduces to discrete peaks, but moreover to a finite number of such peaks, and that this distribution acts as an attractor in any attempt to solve the optimization problem numerically. We also obtain numerical confirmation that $u_n^*(m)=1$ for even $n$, and is different yet similarly positive and non-integrable for odd $n$.

\section{Optimal mass distribution}


\label{sec:num}
\subsection{Numerical methods}
\label{sub:meth}

Three methods were considered to work out the mass dispersion yielding the fastest convergence to
equilibrium, in the discrete or continuous cases. The first two make minimal use of our analytical results, while the last one relies on the developments of section \ref{sec:cont} and consequently proves far more efficient. Yet, all converge toward the same results, as examined in section \ref{sub:res}. In every case, the optimization process has been performed by a hybrid algorithm: at each step, we perform small 
random increments of the optimization variables (the mass distribution), preserving the mean mass and the normalization of the concentration. We then compute the eigenvalue $\lambda_n^{(1)}$ of interest, by diagonalization of a matrix or more efficiently in the third method. If it is lower than its former value, the step is accepted, else it is rejected. The convexity of the landscape ensures that we may apply this simple rejection rule, as there is no local minimum to avoid. Our approach can therefore be
envisioned as a zero temperature Metropolis method, where the largest eigenvalue of a given
matrix plays the role of energy\footnote{We have also tested finite temperature Metropolis algorithms, obtaining similar results.}.

\textbf{Direct resolution:} This approach assumes only that the system is in the vicinity of 
equilibrium, where we may then generalize Eq. 
 \eqref{Udisc} to $N$ species or equivalently express the operator $\mcl K_n$ as a matrix
\beqa &\dt \vec U_n = K_n \,\vec U_n \nonumber\\ 
& ( K_n)_{ij} = \delta_{ij} \sum_{k=1}^N c_k (a^n_{i k} - 1)  + c_j (1-a^2_{ij})^{n/2}& &(\vec U_n)_i \equiv U^n_i(t). 
\label{app1}\eeqa
In order to obtain $\lambda_n^*$, it is therefore necessary to extract the highest eigenvalue $\lambda_n^{(1)}$ of the matrix $ K_n$, and optimize it over the large parameter space $\lbrace c_
i, m_i \rbrace_{i=1,..,N}$ with  $\sum_i c_i=1$ and positivity constraints 
We also wish to check whether our assumption that the optimal distribution must be discrete is correct. A large number of species is used to approximate a continuum. The present method then subdivides into two variants:
\begin{itemize}
\item \textbf{Method A} consists in choosing evenly spaced masses $m_i$, and letting only the concentrations vary, so as to reproduce the outline of the continuous function $c(m)$. Due to invariance of the system by multiplication of all masses by a constant, the only control parameters are the ratio between the 
lowest and highest mass and the number of increments in-between.
\item \textbf{Method B} involves treating each species as a single "particle", with the same concentration $c_i = 1/N$, then letting the masses vary to minimize $\lambda_n^{(1)}$. The distribution $c(m)$ is then computed from the histograms of the number of species having a mass between $m$ and $m+dm$, and the only control parameter is $N$. 
\end{itemize}

\textbf{Resolution by ansatz:} The third approach used is restricted to
even eigen-relaxation modes (even $n$), and takes advantage of the surmise \eqref{saddle} put under matrix form
\beqa  &J \vec c = \lambda_n \vec 1 \\
&J_{ij} = (a^n_{ij} - 1)  + (1-a^2_{ij})^{n/2}   &(\vec c)_i \equiv c_i, \quad(\vec 1)_i = 1, \quad \vec c . \vec 1 = 1.
\eeqa
We proceed as follows: starting from a given mass distribution, which sets the matrix $J$,
we compute $\lambda_n^{(1)} = (J^{-1} \vec 1). \vec 1$. This eigenvalue needs to be optimized only 
over the masses $\lbrace m_i \rbrace_{i=1,..,N}$. Once the optimum is found, the concentration 
profile $c(m)$ follows as a by-product, from $\vec c  = \lambda_n^{(1)} J^{-1} \vec 1$.
This decoupling between masses and concentrations conveniently reduces by half
the dimension of parameter space.

\subsection{Results}
 \label{sub:res}

\begin{figure}
\includegraphics[width=400pt]{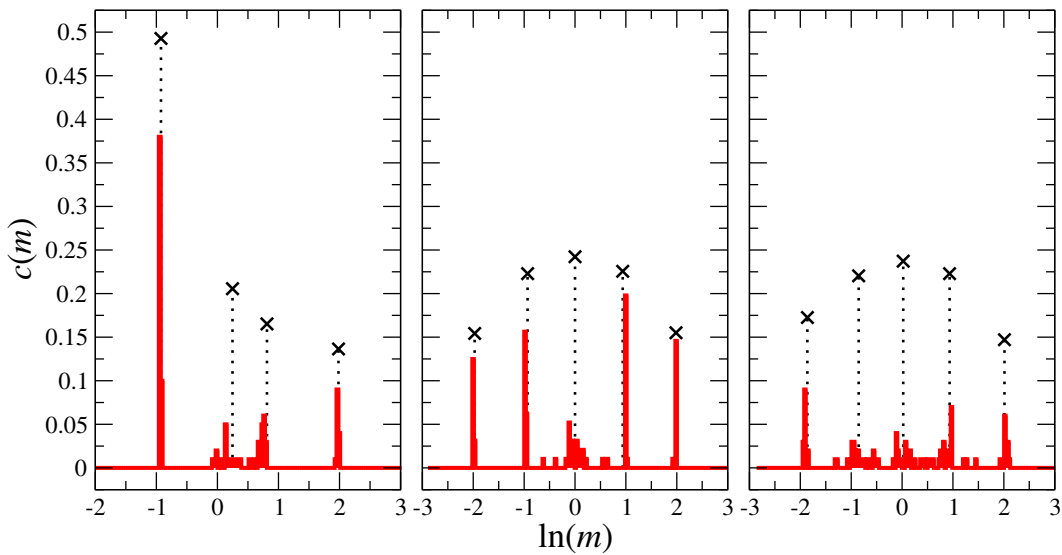}
\caption{ Mass distributions optimized for $\lambda_{3}^{*}$ (left), $\lambda_{4}^{*}$ (center) and the composite eigenvalue $\lambda^*$ (right) corresponding to the true optimum \eqref{eq:deflambdastar} for one-dimensional Maxwell molecules, i.e. $\nu=0$ in equation \eqref{bollin}. 
The crosses represent the optimal mixtures with a few species, either $N=4$ or $N=5$  and the mean mass fixed to unity (their abscissae are reported with dashed lines). The histograms are for a "continuous distribution" approximated with $N=200$ species,  methods A and B giving equivalent results. We see that the continuous case is an approximation of the few-species mixture, and that the optimum for the composite eigenvalue is very close to that of $\lambda_4$ -- although its minimization is more difficult due to its singularity, and leads to a larger spread in the continuous approximant. Note that for $n=4$ (center), $c^*(m)$ is an even function 
of $\log (m)$, up to a trivial shift, which illustrates the invariance under $m \to 1/m$.  }
\label{condist}
\end{figure}

We studied both $\lambda_3^*$ and $\lambda_4^*$ although more emphasis was put 
on the latter. The results were stable under change of initial conditions and optimization method: each degree $n$ seems to be associated with an optimal number of species $N^*(n)$, above which all methods tend to remove surnumerary species by fusing them or making their concentration vanish. Any attempt to prevent this vanishing (for instance adding an energy cost to low concentrations) only creates artificial minima which vary with the regularization method, and the convexity of $\lambda_n^{(1)}[c]$ ensures that if the minimum in the parameter space happens to lie in the subspace where only $N^*$ species have non-vanishing concentrations, then it is also a global minimum for any $N \geq N^*$. 
The study of continuous mass distributions is of particular interest, through methods A and
B with $N\gg 1$. As anticipated, the corresponding optimum distribution is not continuous,
see Fig. \ref{condist} where the histograms exhibit clear gaps in some mass ranges. Furthermore, these
histograms are centered around the support of the discrete solution shown by the c.
However, while the discrete optimum is quickly reached from moderate numbers such as $N=40$, the histograms for the case $N=500$ still exhibit some dispersion, that we attribute to the extreme  flatness of the energy (eigenvalue) landscape close to optimality, see Table \ref{tab1}. 
To substantiate our statement, we have bunched together the histograms in between 
the gaps, see the crosses in Fig. \ref{condist}. The crosses and the circle are quite 
close to each other, and we therefore conclude here that the continuous 
data in the figure are not fully equilibrated, but provide an approximation 
of the correct result. The flatness of the landscape can be appreciated from the
fact that the eigenvalues corresponding to the continuous 
data in Fig. \ref{condist} are, to the 4th digit, the same as their discrete
counterparts reported for $N=4$ (third order moments, $\lambda_3^*$) and $N=5$ 
(fourth order moments, $\lambda_4^*$).

\psfrag{Concentration}{~~~~~~$c(m)$}
\psfrag{Mass}{~~~~~~$m$}

\begin{table}
\begin{tabular}{|c|c|c|c|}
\hline \textbf{Type of mixture} & \textbf{Observed minimum } $\lambda_{3}^{*}$ &  $\lambda_{4}^{*}$&  $\lambda^{*}$\\
\hline Two species & -0.2874 (exact) & -1/4 (exact) & -1/4 (exact) \\
Three species & -0.3070 &-0.2844 (exact) & -0.2841 \\
Four species & -0.3071 &-0.2891 & -0.2888\\
Five species & -0.3071 &-0.2895 & -0.2891\\
``Continuum'' & -0.3071& -0.2895 & -0.2891 \\
\hline
\end{tabular} 
\caption  { Optimal values of $\lambda_{3}^{*}$, $\lambda_{4}^{*}$  and the composite eigenvalue $\lambda^*$ up to the 4th digit for different types of mixtures (numerical results except when otherwise specified). The result for a ``continuous'' dispersion is obtained using both methods A and B described in section \ref{sub:meth}. 
The corresponding mass distributions are shown in Fig.~\ref{condist}:
although the ``continuous'' data still exhibits some spread around the discrete peaks, the associated eigenvalues are extremely close, and cannot be distinguished 
at this precision. Mixtures with $N$ beyond 5 always always yield the same
optimum as the 5 species case (or 4 for $\lambda_3^*$). }
\label{tab1}
\end{table}


 The consistency of these numerical methods lends credence to our conclusions: the minimum $\lambda_3^*$ is found with $N^*=4$ species, and does not change for $N>N^*$, including
for $N\to\infty$ and in the continuous limit. Likewise for
$\lambda_4^*$ with $N^*=5$ species. The relationship $N^*(n)$ for higher moments is not trivial and was not found explicitly, but it is unnecessary for our purposes, as only the $n=3$ and $n=4$ subspaces compete to 
determine the optimal distribution. However, as emphasized above, it is not fully satisfactory to decouple the study of $n=3$ from that of $n=4$ due to crossings in their eigenvalues. It is the reason why we have also considered the composite quantity $\max(\lambda_3^{(1)},\lambda_4^{(1)})$
along the lines of section \ref{sec:binary}. The resulting optimum, the ``true'' one 
for our problem, is also shown in Fig.~\ref{condist}. It can therefore be stated that the generic optimum is made up of five species, and it appears very close to that obtained when restricting minimization to the $n=4$ sub-space.

\subsection{Analysis}
 \label{sub:ana}

 The unexpected result of having a finite mixture as a general optimum may be understood from heuristic considerations: on the one hand, identical particles cannot relax together, and including more species allows for more partners; on the other hand, if these partnerships are to be efficient, each pair of particles needs to have a mass ratio close to the binary optimum, and this constraint is easier to satisfy with a smaller number of species. These opposing effects equilibrate at some $N^*$, whose precise value therefore depends on the specifics of the system. It is also interesting to recall and
illustrate our dynamical equipartition principle for even orders $n$, as 
embodied in Eq. (\ref{saddle}). In the light of this explanation, the significance of the equipartition property becomes clearer: equilibration between species is faster than between particles of the same species, as the latter cannot happen on its own and must be mediated by multiple collisions with heterogeneous partners. Therefore, the optimal path toward equilibrium is the one along which all species first reach the same perturbed distribution, before they all relax together.
And indeed, as remarked in Appendix B, eigenfunctions corresponding to differences between species are associated with more negative eigenvalues, vanishing faster, while the last remaining component of the perturbation is always an additive combination of all species that corresponds to the collective relaxation process.

\begin{figure}
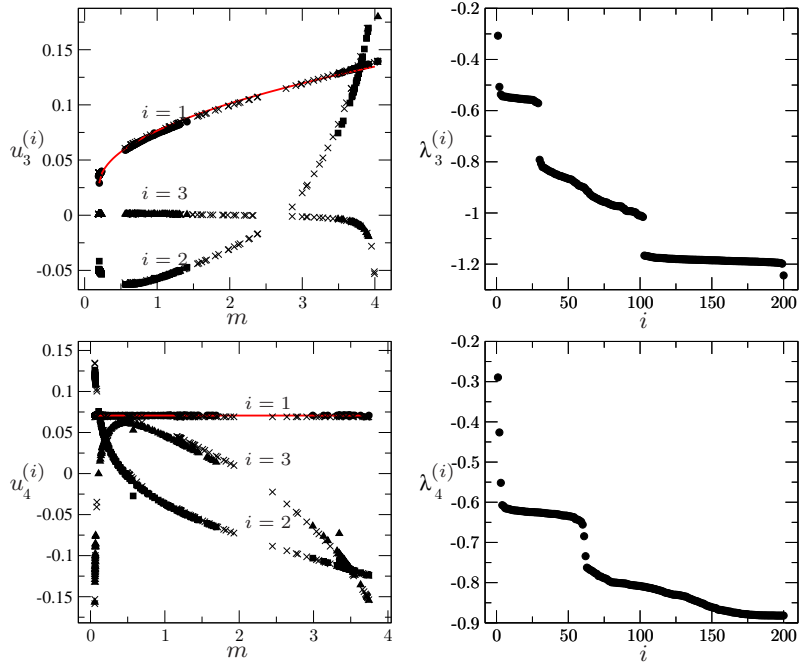

\includegraphics*[width=300pt]{eigensys3.eps}
\includegraphics*[width=300pt]{eigensys4.eps}
\caption{The three  eigenvectors associated with the three largest eigenvalues (left) and the eigenvalue spectrum (right) for $n=3$ (top) and $n=4$ (bottom) for the optimal dispersion $c(m)$ reported in Fig.~\ref{condist}. The index $i$ ranks the eigenvalues in decreasing order. 
Concerning the eigenvectors: crosses are obtained from Method A and full symbols from Method B discussed in section \ref{sub:meth}. No symbols are reported for masses such that $c(m) =0$, hence the gaps in the data. The solid lines correspond to the ansatz $u^{(1)}_3(m)\propto \sqrt{m-m_-}$ (with $m_-$ the minimal mass in the system) and $u^{(1)}_4(m)\propto 1$. Only the second one illustrates the dynamical equipartition 
property alluded to in section \ref{sec:cont}.}
\label{eigsys}
\end{figure}

 An illustration is given by the numerical computation of the optimal eigenfuction $u_n^*(m)$ for different mixtures, which is shown 
in Fig. \ref{eigsys}. The first observation is that indeed, for $n=4$ (and all other even $n$, not displayed here),
the relevant eigenvector $u_4^{(1)}$ is uniform over all masses when the mixture is optimal.
Another feature that emerges from the figure concerns odd moments,
for which the equipartition principle does not hold as such. Instead, the figure shows that
$u_3^{(1)}$, as a function of mass, is very close to a square root law. This shape is in fact not unexpected: it suggests that in writing the perturbation \eqref{eq:defpert}, the relevant physical quantities undergoing dynamical equipartition may not be exactly the odd Hermite moments but rather contain additional prefactors similar to $\sqrt m$. This is readily understood from the first moment, as momentum is given by $mv = \sqrt{m/2} H_1(v\sqrt{m/2})$. Higher odd moments may be interpreted as fluxes, and thus involve momentum as well.  Yet, it was not possible to turn this remark into a consistent ansatz for odd $n$, having the same status as Eq. (\ref{saddle}).

\psfrag{m}{$m$}
\psfrag{i}{$i$}
\psfrag{i=1}{\scriptsize $i=1$}
\psfrag{i=2}{\scriptsize $i=2$}
\psfrag{i=3}{\scriptsize $i=3$}
\psfrag{u}{$u$}
\psfrag{(i)}{\scriptsize ($i$)}
   
\subsection{ Effects of dimensionality }

We may now take interest in higher dimensionality $d>1$, as we wish to ascertain whether we can still find a polydisperse optimum, although relaxation is then possible even in a monodisperse system. In the one-dimensional case, it was shown that considering only symmetric moments and especially the fourth led to a good and more tractable approximation of the general optimization process. It is therefore reasonable to focus here likewise on the fourth isotropic moment, restricting the perturbation to
\beq F_i(\bv,t)= c_i \,\mcl M(m_i,\bv) \, \left[ 1+U^4_i(t) S_2(\bv^2 \sqrt{m_i/2}) \right]\eeq
where the $d$-dimensional  Sonine polynomials appear as a generalization of the Hermite polynomials and may be written as associated Laguerre polynomials $L_n^p(x)$ \cite{arfken2005mathematical}
\[S_n(x)=\dfrac{1}{n!} e^x x^{1-d/2} \dfrac{d^n}{d^n} \left(  e^{-x} x^{n-1+d/2}\right) = L_n^{d/2-1}(x)\]
The matrix for the fourth moment is then given by
\[d(d+2) (K_4)_{ij}=- \delta_{ij}\left[2(d-1)c_i + \sum_{k\neq i} \left(2d+1 + 2(1-d) a_{ij}^2-3 a_{ik}^4\right) c_k \right] + (1-\delta_{ij}) 3(1-a_{ij}^2)^2 c_j.\]
This expression may be inserted in the same optimization procedures as for one-dimensional mixtures. The results  are reported in Table \ref{tab2}, where two important observations can be made. First, the optimum is always attached to a discrete mixture with a finite number of species, which however decreases as the dimensionality increases. 
\begin{table}
\begin{tabular}{|c|c|c|c|}
\hline \textbf{Dimensionality} & \textbf{Optimal relaxation rate} & \textbf{Number of species $N^*$} & \textbf{Parameters}\\
\hline 1 &  -0.2874 & 5 & see Fig.~\ref{condist}\\
\hline 2 & $-0.2922$ & 3 & $m_2/m_1\approx 2.32 $, $c_1\approx 0.35$\\
\hline 3 & $-49/180 \approx -0.2722$ & 2 & $m_2/m_1 \approx 2.38$, $c_1=c_2$ \\
\hline 4 & $- 1/4$ &1& \\
\hline
\end{tabular} 
\caption{Optimal mixture for the relaxation of the fourth Sonine moment depending on the dimension. For $d=2$, note that $m_3/m_2=m_2/m_1$ and $c_3=c_1$ due to the symmetries of the evolution operator. For $d=3$ see the analytic expression in Fig.~\ref{fig:binnd}. In all cases, the optimal eigenvector verifies the dynamic equipartition principle: all its components are equal.}
\label{tab2}
\end{table}
Second, in more than three dimensions, the optimal mixture reduces to a single species, as can be explicitly checked from the expression of the upper eigenvalue for the binary mixture
\[\lambda_4^+ = \dfrac{(1-d)(1+\zeta^2)^2 +12 (3-d) \zeta^2 -4(1+2d)(1+\zeta^2)\zeta} {d (  d+2)(1 + \zeta)^4} \]
which ceases to have extrema at $m_2/m_1 \neq 1$ for $d\geq 4$. Therefore, we find an upper critical dimension of sorts at $d=4$, defining the threshold for optimality of the monodisperse system.
The most physically relevant choice $d=3$ corresponds to $N^*=2$, and a simple binary mixture therefore allows for optimal relaxation -- however the effect of mass heterogeneity is minimal: a $2$ percent increase of the relaxation rate. Thus, optimal polydispersity is much more significant in monolayer (effectively bidimensional) flows, where the corresponding increase is found in excess of 20 percent. Also, a spread in particle sizes still has to be considered and could prove more relevant in higher dimensions.

\begin{figure}
\begin{tabular}{m{6cm}m{6cm}}
\includegraphics*[width=180pt]{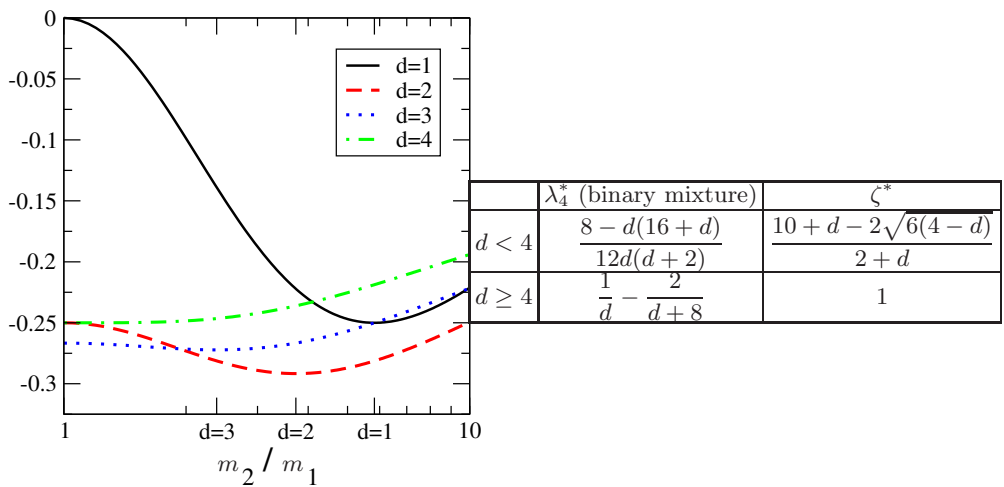}& \begin{tabular}{|c|c|c|}\hline
 & $\lambda_{4}^*$ (binary mixture)& $\zeta^*$ \\ \hline
$d<4$& $\dfrac{8-d(16+d)}{12d(d+2)}$&$\dfrac{10+d-2\sqrt{6(4-d)}}{2+d}$\\ \hline
$d\geq 4$& $\dfrac{1}{d}-\dfrac{2}{d+8}$&1\\ \hline
\end{tabular}
\end{tabular}
\caption{\label{fig:binnd} Left: Relaxation rate for the $d$-dimensional Sonine moment $\langle S_n(mv^2/2)\rangle$ of an equimolar binary mixture, in dimension $d=1$ up to $4$. The locations of the optimal mass ratios are marked on the $x$-axis for $d<4$: although relaxation is possible in the monodisperse case $\zeta = 1$ for any $d>1$, there remains a polydisperse optimum until $d=4$. Right: Optimal relaxation rate and corresponding mass ratio for a binary mixture in arbitrary dimension $d$ (it is necessarily equimolar due to the symmetries of the operator). }\end{figure}

\section{Beyond the elastic Maxwell model}
\label{sec:ext}

We have so far concentrated on Maxwell molecules ($\nu=0$)
undergoing energy conserving
collisions. We generalize the approach to different collision kernels, allowing
also for dissipation upon binary encounters. 

\subsection{Other collision kernels}

\begin{figure}
\begin{center}
\includegraphics*[height=160pt]{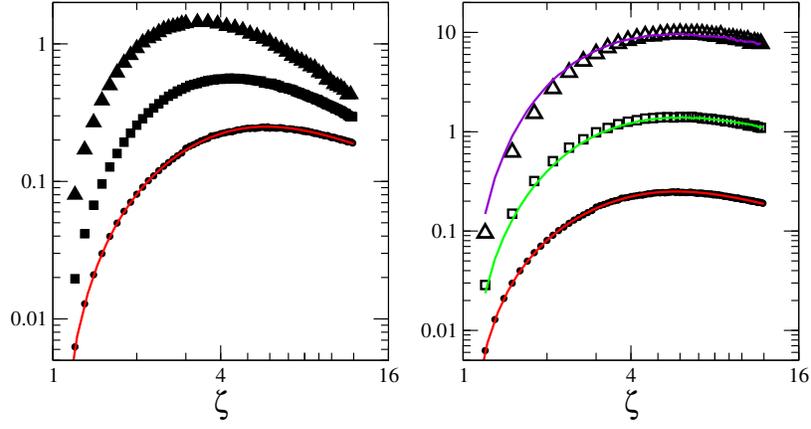}
\caption{ Relaxation rate of the fourth moment for different collisions kernels $\nu=0,1,2$ (circles, squares, triangles) in an equimolar binary mixture, fixing either the velocity scale (left) or the temperature and mean mass (right). For Maxwellian particles $\nu=0$, both constraints give the same result $|\lambda_4^+|$ (solid line, left picture). However, for hard or very hard particles ($\nu=1,2$) the choice of constraints displaces the maximum through a change of scale: the data in the first picture are multiplied by $m^{\nu/2} 1.2^\nu$ to give the solid lines in the second (the numeric factor is adjusted by fit). The constraint of fixed temperature and mean mass leads to all three models exhibiting optimal mass ratios within ten percent of each other.  \label{vhpdim}}
\end{center}

\end{figure}

Both the exact diagonalization of the linearized system described in section \ref{sub:lin} and the Fourier transform technique discussed in Appendix A cease to be applicable when $\nu \neq 0$, as a third term containing the relative velocity of the collision partners enters into the integral in the Boltzmann equation \eqref{bolsys}. Numerical results may nevertheless be obtained through Monte Carlo integration of the equation, employing the widespread Direct Simulation Monte Carlo (DSMC) technique and especially its variant known as Bird's algorithm \cite{Bird}. One should however pay attention beforehand in setting some optimization constraints, in order to have a well posed problem: when one makes the Boltzmann equation dimensionless, the time scale is found to vary as $t\sim v_{12}^\nu \sim \left(\Theta / m\right)^{\nu/2}$
with $\Theta$ the temperature, $m$ a mass scale, and $v_{12}$ the typical scale for the relative velocity of the collision partners. Hence, if we fix $\Theta$ and let the masses vary, the time scale will change as $m^{-\nu/2}$, altering the relaxation rate which we are trying to optimize over mass distributions, except of course for the Maxwell model $\nu=0$. There are two natural choices of constraints, which lead to different optima (connected through a rescaling) as seen in Fig.~\ref{vhpdim}: either we fix the velocity scale $v_{12}$, in which case the temperature of the mixture will vary with the masses, or we fix both the temperature and the mean (or total) mass in the system.  Each choice will conduce to determining the fastest-relaxing mixture in a given context:  the first one is relevant if the  initial velocities are imposed regardless of the mass of the particles (giving more energy to more massive species) for instance with a piston, while the second corresponds to an unbiased initial energy distribution and looking for the optimal partition of a predefined total mass. 

Some comparison is given with the result of the DSMC algorithms in Fig.~\ref{vhpdim}. The long-time decay rate of the  fourth moment is measured for each species' velocity distribution with various mass ratios.
Once the time evolution of this moment has reached a clear exponential behaviour, it can be measured 
and then compared to the analytically determined value of $\lambda_3^+$ in the case $d=1,\nu=0$. 
It can be seen in Fig. \ref{vhpdim} that the phenomenology is quite the same as with $\nu=0$.
Moreover, taking the fixed temperature and mean mass route, the optimum mass ratio
is virtually independent on the value of the kernel exponent $\nu$: 
Maxwellian, hard and very hard particles behave the same, and all other results derived in this paper for $\nu=0$ are expected to hold for realistic hard spheres. 
This remark is fully consistent with previous studies that concluded to 
the usefulness and sometimes accuracy of the Maxwell approach in 
conservative fluids, see e.g. \cite{GaSaBook} for an overview.

\subsection{Inelastic binary Maxwell mixture}
\label{sub:inel}

Finally, one may be interested in ascertaining whether an optimality property is retained if we let the collisions become dissipative. This prevents the system from reaching any equilibrium save for its ultimate state of rest once the initial energy has completely vanished, i.e. $F_i(v,t) \to c_i \delta(v)$ as $t\to \infty$. However a homogeneous gas of inelastic Maxwell particles (see e.g. \cite{GaSa11} for a review) is known to rapidly enter a phase where its decay takes the form of a scaling solution \cite{Baldassarri}, and it may be interesting to quantify the time of approach toward that solution\footnote{Let us remark that, for a monodisperse inelastic 1D system, the shape of the scaling solution is not universal due to a non-vanishing rescaled heat flux \cite{GaSa07}; yet we see that this problem disappears for a mixture.}. Another possibility is to add an external forcing that constantly injects energy into the system, so that it may reach a non-equilibrium stationary state where energy gain and loss compensate. The investigation presented below is but a sketch of a more exhaustive work that we leave to further inquiries, yet it is sufficient for our purpose: it shows that the notion of optimality, and the methods we developed to establish it, do not always endure through the addition of inelasticity, while some remnants of the properties of the elastic system may be found in a region of the parameter space.

Inelasticity is represented in the equations by a very simple redefinition of coefficient $a_{ij}$, 
\beq a_{ij} =\dfrac{m_i-\alpha_{ij} m_j}{m_i + m_j}\eeq
introducing the restitution parameter $\alpha_{ij} \in [0,1]$,  $1$ amounting to elastic collisions between species $i$ and $j$ and $0$ to completely inelastic collisions \cite{BrPoBook}. The simplest case is that of uniform inelasticity $\alpha_{ij} = \alpha < 1$. Then $a_{i i} \neq 0$ and collision of same-species particles contribute to the evolution of the probability. Furthermore, following the derivation in \cite{vNETP99}, we introduce the external forcing under the guise of a white noise characterized by a coefficient $D$. Finally, we take advantage of the convolution structure of the Boltzmann equation for $\nu=0$, which allows for a simplification in Fourier space as reported in Appendix A:
\beqa 
(\dt+1+D_i k^2 ) \widehat F_i(k,t)    =c_i \,\widehat  F_i(a_{ii}k,t)\,\widehat F_i((a_{ii}-1)k,t) +  c_j \,\widehat  F_i(a_{i  j}k,t)\,\widehat F_j((a_{ij}-1)k,t). \label{difin} 
\eeqa 
where we used the shorthand $D_i = D/m_i^2$. We may then employ the moment expansion method by performing a Taylor expansion in $k$
\beq\widehat F_i (k,t) = \sum_{n=0}^\infty \dfrac{(ik)^n}{n!} \mu_i^n(t), \eeq
and the connection with our previous approach appears as the coefficients $\mu_i^n(t)$ are in fact the moments of the distribution $F_i(v,t)$
\beq\mu_i^n(t) =\int_\RR dv \,v^n \,F_i(v,t).\eeq
We thus obtain the general expression 
\beq (\dt+1) \mu_i^n(t) =D_i n(n-1) \mu_i^{n-2}(t) + \sum_{m=0}^n \dbinom{n}{m} \mu_i^{n-m}(t) \left[ c_i a_{ii}^{n-m} (a_{ii}-1)^{m}  \mu_i^m(t) +   c_j a_{ij}^{n-m} (a_{ij}-1)^{m}  \mu_j^m(t)\right] \eeq
The role of the bath in altering the long-time behavior of the system is made apparent in this hierarchy: $D=0$ allows the trivial stationary solution $\mu_i^n = 0$  for all $n >0$ while any $D \neq 0 $ excludes this solution since higher moments then depend linearly on $\mu_i^0$, which cannot vanish as it is the normalization of $F_i(v,t)$.

Considering centered initial distributions, $\mu_i^1(t)=0$, the equations for the second and third moments take a simple form reminiscent of \eqref{app1}
\beqa &(\dt+1) \vec \mu_n =2 D_i  \delta_{n,2}+  M_n \vec \mu_n \quad n=2,3 \nonumber \\ 
& ( M_n)_{ij} = \delta_{ij} \sum_{k=1}^2 c_k a^n_{i k}  + c_j(1-a_{ij})^n& &(\vec \mu_n)_i \equiv \mu^n_i(t). 
\label{Winel}\eeqa
from which we easily recover stationary values $\mu_i^2(\infty)\propto D_i$ and $\mu_i^3(\infty)=0$ as well as the eigenvalues $\lambda_n^\pm$ of $ M_n$
\beq \lambda_n^\pm = \dfrac{1}{2}\left(\Tr  M_n \pm \sqrt{\Tr^2  M_n - \Det M_n} \right) \eeq
which describe the approach toward these values. Higher moments will have similarly defined eigenvalues, as well as exponential forcing from the lower degrees adding all combinations 
$\lambda_j^\pm+\lambda_k^\pm$ with $j+k=n$ (provided $j,k\geq 2$) and $\lambda_{n-2}^\pm$ coming from the bath term, among which the maximal eigenvalue must be found and then minimized to optimize the relaxation rate of the moment. In practice, the noise amplitude $D$ may depend on the masses
(some arguments along these lines were given in \cite{BaTr02GM}), but we discarded
such an effect for simplicity.

\begin{figure}
\centering
\includegraphics*[width=160pt]{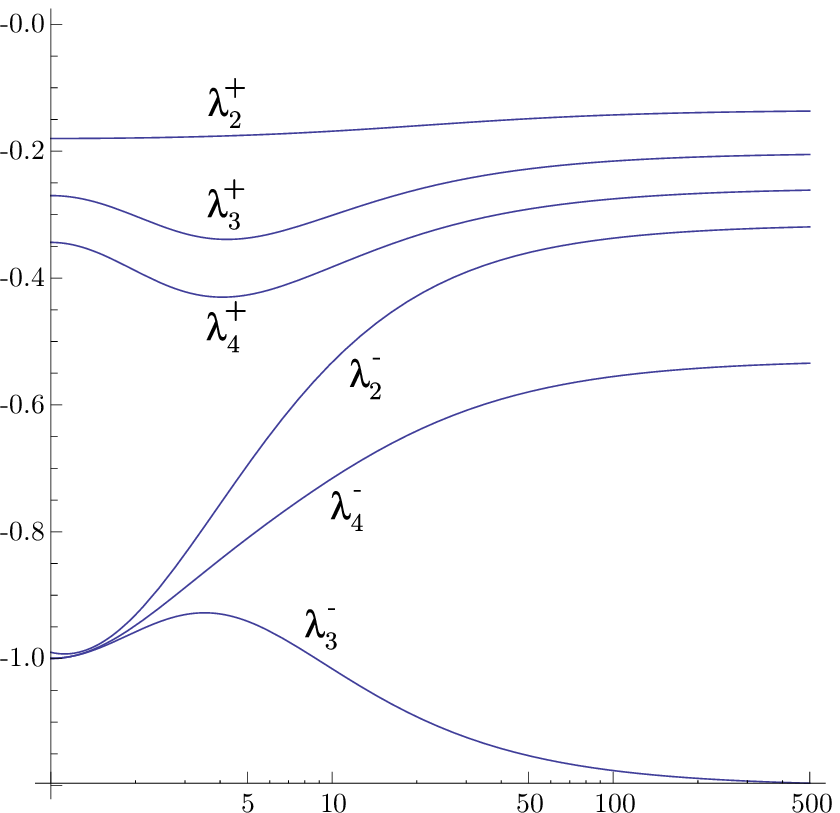}\includegraphics*[width=160pt]{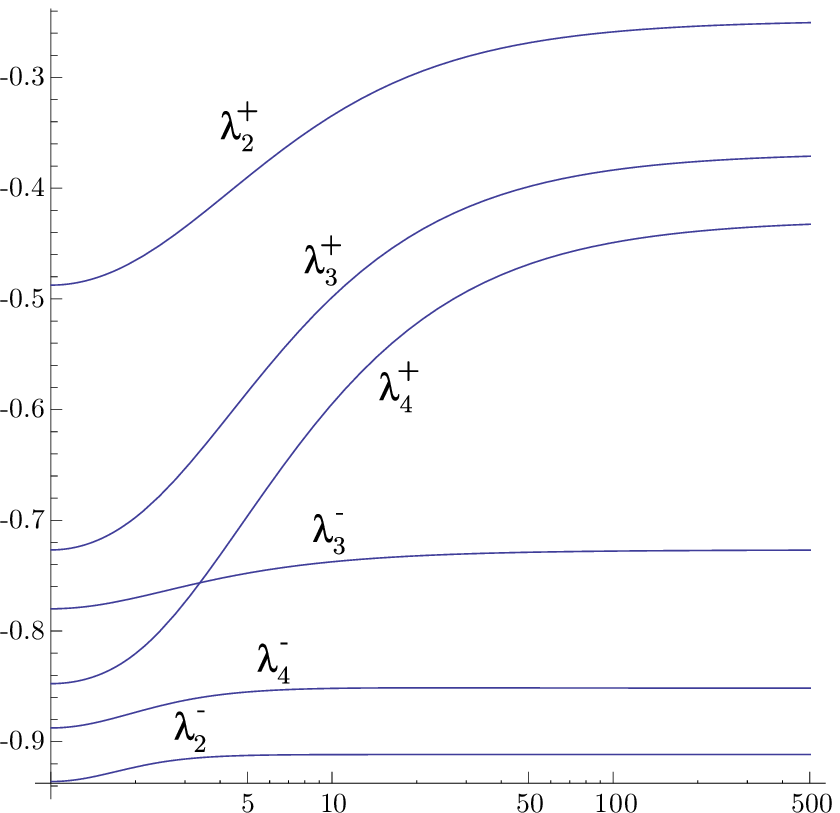}
\includegraphics*[width=160pt]{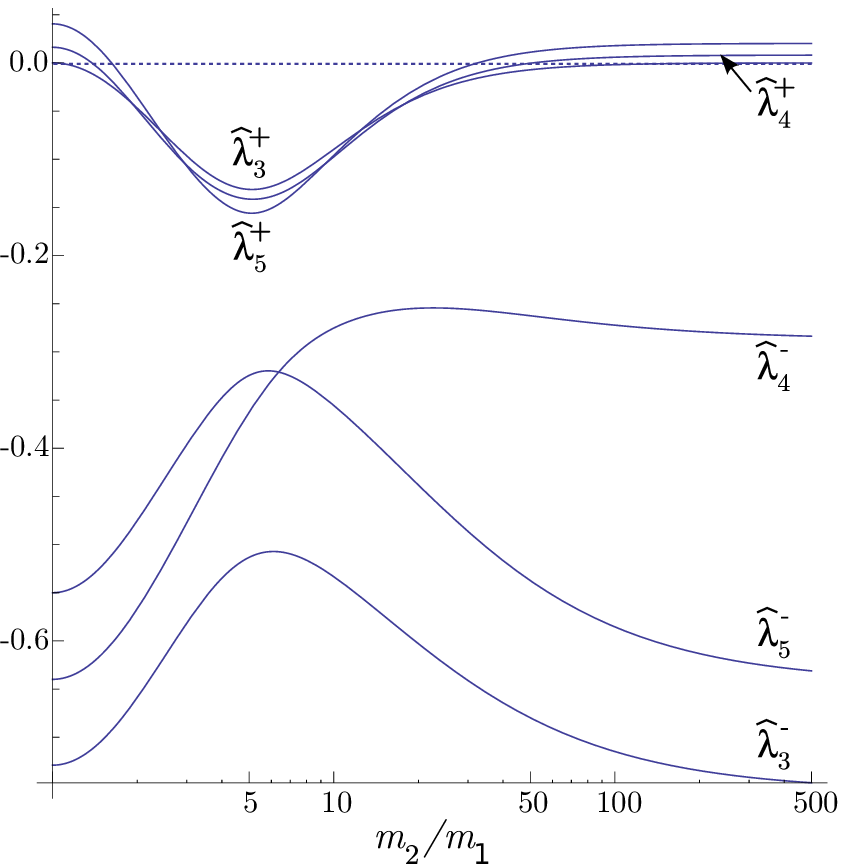}\includegraphics*[width=160pt]{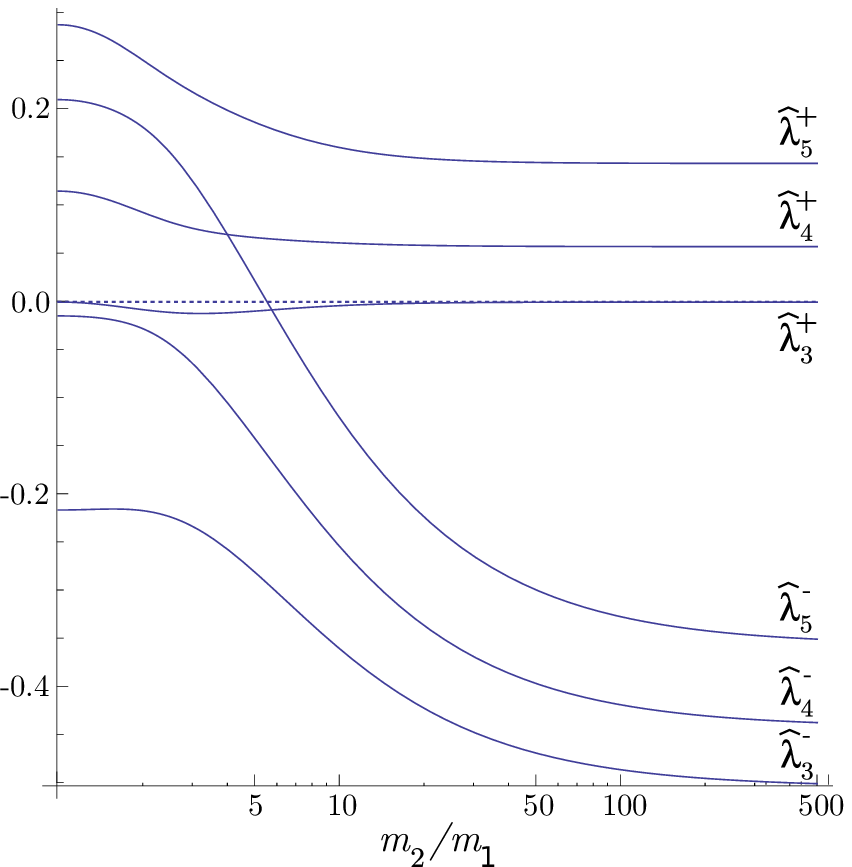}
\caption{Eigenvalues spectrum for the inelastic equimolar binary mixture versus mass ratio, with $\alpha= 0.8$ (left) and $\alpha =0.2$ (right). Top: spectrum in the forced state. Bottom: rescaled eigenvalues in the free cooling state (the dotted line materializes the ordinate $0$ allowing to see all eigenvalues that cross it, representing diverging moments). \label{fig:spectrdiss}}

\end{figure}

\begin{figure}
\includegraphics*[height=140pt,clip]{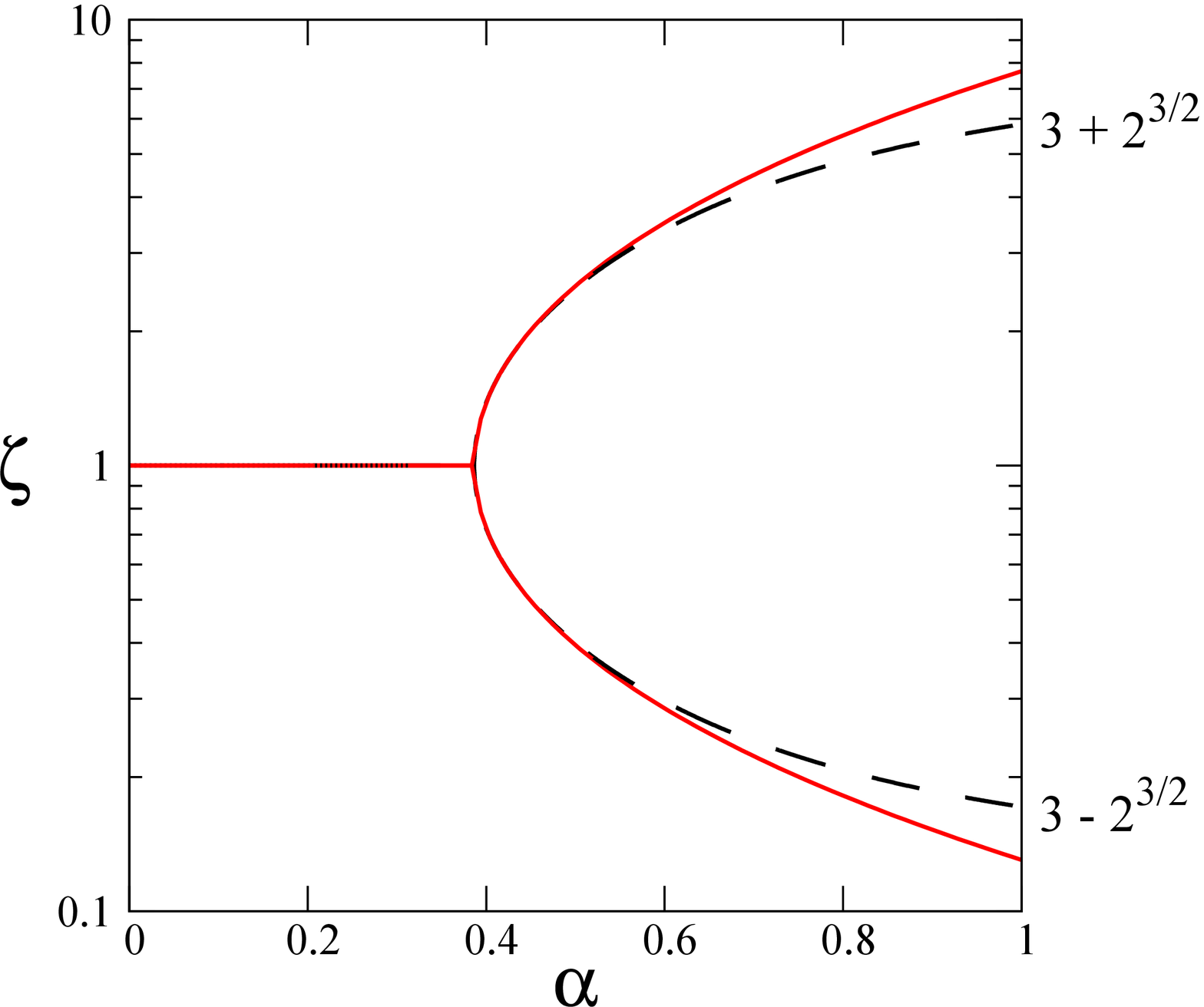}\includegraphics*[height=140pt,clip]{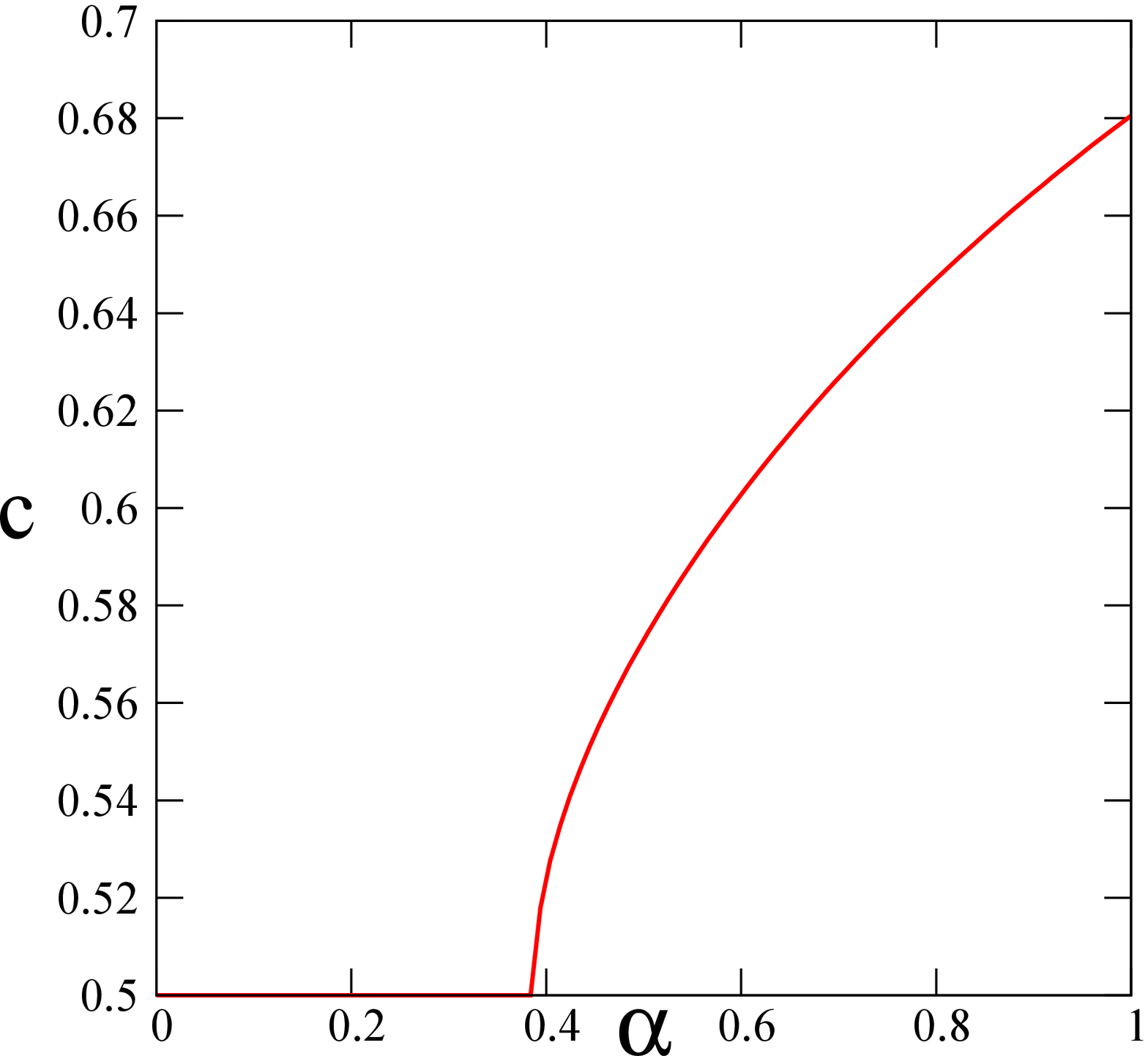}
\caption{Optimization of  $\lambda_{3}^{+}$ in the (forced) binary mixture with finite restitution coefficient $\alpha \in[0,1]$. The dashed branch ending at $3\pm 2\sqrt{2}$ shows the equimolar
result, while the more general solution allowing for unequal concentrations is shown with
the continuous curve. 
The corresponding values of $\zeta=m_2/m_1$ and $c_1$ are found to vary continuously from the elastic limit $\alpha = 1$ to the threshold $\alpha_c \approx 0.37$, where the optimum becomes the single-component gas $m_2 = m_1$. This bifurcation does not appear for the rescaled eigenvalues of the free cooling state.}
\label{inelastic}
\end{figure}

Two quite different results are obtained whether one retains the external forcing or not, see the spectra 
in Fig. \ref{fig:spectrdiss}. In the forced case, the highest eigenvalue is always $\lambda_2^+$ which is non-zero (contrary to an elastic gas) as it describes the approach toward the non equilibrium steady state where forcing and dissipation compensate. It is minimal for equal masses, and there is no optimal non-trivial polydispersity in that case. If however the forcing is adjusted so that the initial energy is conserved, this mode does not appear (except for the much faster-relaxing component $\lambda_2^-$ leading to unequal granular temperatures $ m_i \mu_i^2(\infty)/2$). We may then look for the optimum of other eigenvalues such as $\lambda_3^+$, find their expression for all $\alpha$, and subsequently compute the optimal mass ratio and concentrations. As seen in Fig.~\ref{inelastic}, the optimum for $\lambda_3^{+}$ presents a bifurcation for a finite value of $\alpha$: under the threshold $\alpha_c \approx 0.37$, the third moment vanishes fastest for identical masses and any mixture is suboptimal. Interestingly, although the form of the solution $F_i(v,t)$ is highly singular in the elastic limit $\alpha \to 1$, the eigenvalue $\lambda_3^*$ is found to behave regularly and the optimal parameter values all vary continuously for $\alpha \in [0,1]$.

If however we consider the freely cooling solution with $D=0$, we are more interested in the approach to the scaling solution
\beq F_i(v,t) = F_i\left(\dfrac{v}{v_0(t)} \right) \approx  \dfrac{ 1}{\left(1 + \left(v/v_0 \right)^2  \right)^{\sigma} }  \label{scalingsol} \eeq
(see the derivation in \cite{Marconi}) with $v_0(t)=\sqrt{\mu_i^2(t)}$ the decreasing thermal velocity and $\sigma \geq 2$ an exponent depending on $\alpha$ and $m_1/m_2$.

As we may infer from the power-law tails of the scaling solution \eqref{scalingsol}, higher-order moments diverge when computed over rescaled velocities $v/v_0(t)$ \cite{Marconi,GaSa07}: in this frame, the solution is stationary but the eigenvalues associated with the rescaled moments become 
\beq \widehat \lambda_n^\pm = \lambda_n^\pm - \dfrac{n}{2} \lambda_2^+.
\eeq 

Depending on the parameter values, we may have $\widehat \lambda_n^+ > 0$, representing the diverging moments.  Optimal convergence toward the scaling solution is plausibly best described by the new criterion
\beq \widehat \lambda^* = \max_{c_i,m_i} \min_{n,j} |\lambda^{(j)}_n|\eeq
corresponding to the fastest evolution of the slowest mode, regardless of whether it is converging or diverging. This criterion is rather difficult to manipulate as there are now many eigenvalue crossings, but we may still select $\widehat \lambda_3^+$ as it seems relevant in both the  quasi elastic and totally inelastic limits. The mass ratio for an equimolar binary mixture that optimizes relaxation of the rescaled third moment is given by
\beq \zeta \approx 5.8 \quad (\alpha \approx 1) \qquad \qquad \zeta \to 2 \quad (\alpha \to 0). \eeq
It is thus reasonable to expect that, contrary to the forced case, a polydisperse optimum exists for all values of the inelasticity. In both cases, it extends continuously from the elastic optimum, and therefore we may at least state that our results for conservative systems remain mostly relevant at small inelasticities.
The extension to multiple species may also be attempted: since the introduction of inelasticity results in a contraction of the binary optimum as observed for the elastic mixture in higher dimensions, we expect likewise that the optimal number of species in the mixture will decrease with the restitution coefficient.


\section{Conclusions}

Although polydispersity is an important property of systems with mesoscopic components such as colloidal or granular fluids, and is known to induce many non-trivial behaviours, it is seldom considered from the perspective of optimization. Our aim in this paper was to ascertain whether one could find some optimal dispersion with respect to a simple property, the relaxation time of a slightly perturbed gas near equilibrium. We conclude that such optimum can be found, but it is a finite discrete mixture rather than a continuous dispersion. The ideal number of species depends on the dimension of space, and reduces to one for $d\geq 4$.

We also observe that relaxation in a mixture is a two-step process, tending to equalize the velocity distributions of different species long before each one reaches its asymptotic form (be it equilibrium or a stationary out-of-equilibrium state).  Indeed, the last quantity to relax is always an additive combination of moments for all species, rather than some difference between them. Therefore, such observables as the temperature ratio of two species do not properly allow to check whether the system has truly relaxed ; one ought rather to measure the evolution of some global quantity such as the sum of fourth moments. Furthermore, relaxation is optimal when the composition of the mixture allows the first-step equalization process to be maximally efficient, leading to a dynamical equipartition of the perturbation between all species.

These results, although computed for Maxwell particles, remain  valid for hard particles, as we show in the simple binary case that a change of collision kernel has minimal consequences on the behavior of the system.  The evolution of moments in a binary mixture of inelastic Maxwell particles has been demonstrated to exhibit some divergences \cite{Marconi}, therefore only limited correspondences may be drawn with the universal tendency of return to equilibrium that is found in elastic gases; yet, we can look for mixtures that exhibit optimal evolution toward some setting-specific asymptotic state and it appears that these optima are simply continuously displaced from equilibrium results if inelasticity is small enough. Natural extensions of this work include investigating 
optimality conditions with respect to more experimentally relevant properties such as transport coefficients.

\bibliography{continu2}

\section*{Acknowledgments}
We would like to thank G. Allaire, E. Bogomolny, O. Giraud, P.-E. Larré, Y. Atas and D. Villamaina for meaningful discussions.

\section*{Appendix A: Nonlinear equations for a specific initial condition}
\label{sub:nlin}

 We report here the second method that was proposed for the elastic maxwellian binary mixture \cite{Mohazzabi,Mohaz2,Mohaz3}, as it leads to the same optimal mass ratio \eqref{stdratio} as found by the previous approach. We wish to demonstrate the connection between both methods, so as to further point out the adequacy of our first definition for the relaxation rate. The second approach avoids the linearization process through the rewriting of the equations under a form that may be solved for a specific initial condition.

 The right-hand side of \eqref{bolsys} with $\nu=0$ may be seen as a convolution, and it is natural to take the Fourier transform $\widehat F(k,t) = \int_\RR dv \,e^{-ikv}\, F(v,t)$ to convert this convolution into a simple product 
\beqa(\dt+1) \widehat F_i(k,t) =\widehat  F_i(a_{ij}k,t)\,\widehat F_j((a_{ij}-1)k,t). \label{diffunc}\eeqa

An exact solution of \eqref{diffunc} can also be obtained with additional simplifications. Discretizing time, one may replace $(\dt+1)\widehat  F_1(k,t)$ by $\widehat F_1(k,t+1)$, and rewrite \eqref{diffunc} as a difference-functional equation system. The chosen initial condition is bimodal for one species while the other is at rest
\beqa F_1(v,0)&=\frac{1}{2}\left(\delta(v-v_0) + \delta(v+v_0)\right) \nonumber \\
F_2(v,0)& =\delta(v). \eeqa
The solution \cite{Mohazzabi,Mohaz2,Mohaz3} will be reported here for species $1$ only, that may be expressed for even times $t=2j$ as
\beq\widehat F_1(k,2j) = \prod_{m=0}^j \cos^{\scriptstyle 
\left(\begin{matrix} \scriptstyle 
2j \\ \scriptstyle 2m \end{matrix}\right)}\left[a^{2j-2m} \,(a^2-1)^m \,k \right]\eeq
where the cosine is elevated to the power given by the number of combinations of $2m$ elements out of $2j$. For sufficiently large $j$, this leads to the approximation
\beq\widehat F_1(k,2j) \approx \exp\left[-(k^2/4)(1+\xi^{2j})\right] \eeq
with 
\beq\xi = \dfrac{m_1^2 - 6\, m_1 m_2 + m_2^2}{(m_1+m_2)^2} \eeq
a coefficient previously seen in the heuristic argument for \eqref{stdratio}. The connection of this result with the previous method is made explicit by taking the Taylor expansion of the approximation above
\beq \widehat F_1(k,2j) \approx  \exp\left[-\dfrac{k^2}4\right]  \sum_{n=0}^\infty \dfrac{1}{n!} \left(-\dfrac{k^2}{4} \xi^{2j}\right)^n \eeq
which becomes, under inverse Fourier transformation
\beq F_1(v,2j) \approx e^{-v^2} \sum_{n=0}^\infty \dfrac{1}{4^n n!} H_{2n}(v) \,(\xi^{2n})^j.\eeq
Hence, with a rescaling of the reference temperature, we may relate the linearization method to the special solution given here, in the long time limit $j\gg1$ which is the condition of validity for the expressions above. In that limit, the system has already relaxed if $\xi = 0$, while it is still out-of-equilibrium for any other value of $\xi$.  It is clear therefore that the optimal mass ratio must be given by the roots of $\xi$, that is $m_1/m_2 = 3 \pm 2\sqrt 2$. This result was confirmed by numerical simulation of Maxwell particles on a line in \cite{Mohazzabi,Mohaz2,Mohaz3}.

\section*{Appendix B: Properties of the optimal eigenfunction}
Let us assume that some eigenvalue $\lambda_n$ is associated with the constant eigenfunction $u_n(m)=1$. Then according to the definition of $\mcl \mcl K_n$ in \eqref{eq:Ucont} the corresponding mass distribution $c$ must verify
\beq \lambda_n = \int_\Dm \! dm' \ c(m') \left[ a^{n}(m,m') -1 +  \,  \left(1-a^2(m,m') \right)^{\frac{n}{2}} \right] \quad \forall m \in \Dm\label{eigbare}\eeq
Our conjecture is that this choice of eigenfunction $u_n$, and the distribution $c_n$ defined implicitly as the one minimizing $\lambda_n$ in Eq. \eqref{eigbare}, give the sought saddle point for even $n$. We have no rigorous proof of this conjecture, which is nonetheless sustained by induction: both from analytical expressions in the case of a mixture with few species, where the optimal eigenvector for even moments is always the "constant" vector with all its components equal, and from numerical observations presented in section \ref{sub:res}.

Nevertheless, the most physically relevant property of this optimal eigenfunction can be understood from a weaker result that we show here: for any $n$, any eigenfunction that has no zero on $\Dm$ is necessarily the highest eigenfunction for the corresponding dispersion $c(m)$. 
The justfication of the statement is as follows: we start from 
\beq \lambda_n = \int_\Dm \! dm' \ c(m') \left[ a^{n}(m,m') -1 +  \dfrac{u_n(m')}{u_n(m)}\,  \left(1-a^2(m,m') \right)^{\frac{n}{2}} \right] \eeq
For any choice of $c$ verifying this equation, and for any function $u$ we may write the scalar product
\beq\langle \mcl K_n u, u\rangle =\int_\Dm \! dm \,c(m)\, u(m) \int_\Dm \! dm' \ c(m') \left[ \, u(m) \, (a^{n} -1) +  u(m') \,  (1-a^2 )^{\frac{n}{2}} \right]  \eeq
then use the substitution involving this particular non-vanishing eigenfunction  $u_n(m)$
\beq\int_\Dm \! dm' \ c(m') \left( a^{n} -1\right)  = \lambda_n - \int_\Dm \! dm' \ c(m')\,\dfrac{u_n(m')}{u_n(m)} \left(1-a^2 \right)^{\frac{n}{2}}   \eeq
to obtain
\beq\langle \mcl K_n u, u\rangle =\lambda_n \langle u, u\rangle + \int_{0}^{\infty} \! dm\, dm' \,c(m)\, c(m')   \,u(m) \left[u(m')  - u(m)\dfrac{u_n(m')}{u_n(m)} \right]  (1-a^2 )^{\frac{n}{2}} .  \eeq
Now, the integral term is easily shown to be non-positive: permuting the integration variables and taking the half-sum gives
\beq-\dfrac{1}{2}\int_\Dm \! dm\, dm' \,\dfrac{c(m)\, c(m')}{u_n(m) \, u_n(m')}   [u(m)-u(m')]^2  (1-a^2 )^{\frac{n}{2}}  \leq 0\eeq
all terms inside the integral being non-negative, so that
\beq\langle \mcl K_n u, u\rangle \leq \lambda_n \langle u, u\rangle \quad \forall u. \eeq
Hence, $\lambda_n = \lambda_n^{(1)}[c]$ is the maximal eigenvalue associated with distribution $c$, with $u_n = u_n^{(1)}[c]$ the corresponding eigenfunction. By this reasoning, one may obtain similar conclusions for any symmetric operator defined as the sum of a diagonal term and an operator with non-negative kernel: in any such case, an eigenfunction that never vanishes is necessarily associated with the highest eigenvalue.

Furthermore, the eigenvectors of $\mcl K_n$ form an orthonormal basis as it is a Hilbert-Schmidt operator and we have symmetrized it by including the measure $c(m)$, therefore any two eigenfuction $u^{(i)}_n$, $u^{(j)}_n $, $j\neq i$ have vanishing scalar product
\beq\langle u^{(i)}_n ,u^{(j)}_n \rangle = \int_\Dm \! dm \,c(m)\, u^{(i)}_n(m) \,u^{(j)}_n(m) \propto \delta_{ij}\eeq
and any eigenfunction associated with some $\lambda_n^{(i)}< \lambda_n^*$ must have zeros on $\Dm$ for its scalar product with the function $u_n^*$ and measure $c(m)$ (both of constant sign) to vanish. This property is crucial for the physical relevance of our observations: it means that "asymmetric" (zero mass) combinations of moments of multiple species always relax faster than the single "symmetric" or additive combination $u_n^*$. This implies that the velocity distributions of all species become similar before relaxing all together toward equilibrium. Equilibration between species is thus an improper measure of relaxation as it occurs much faster than the rest of the process.

\end{document}